\documentclass[english,amssymb,floatfix]{revtex4}
\usepackage[T1]{fontenc}
\usepackage[latin9]{inputenc}
\usepackage{amsmath}
\usepackage{amssymb}
\usepackage{graphicx}
\usepackage{esint}

\makeatletter

\@ifundefined{textcolor}{}
{%
 \definecolor{BLACK}{gray}{0}
 \definecolor{WHITE}{gray}{1}
 \definecolor{RED}{rgb}{1,0,0}
 \definecolor{GREEN}{rgb}{0,1,0}
 \definecolor{BLUE}{rgb}{0,0,1}
 \definecolor{CYAN}{cmyk}{1,0,0,0}
 \definecolor{MAGENTA}{cmyk}{0,1,0,0}
 \definecolor{YELLOW}{cmyk}{0,0,1,0}
}

\usepackage{amsfonts}\usepackage{bbm}\usepackage{epsfig}\usepackage{graphics}
\usepackage{ifpdf}\ifpdf
\DeclareGraphicsExtensions{.pdf,.png}\usepackage{pgf}\usepackage{tikz}\usepackage{epstopdf}

\else
\DeclareGraphicsExtensions{.eps,.png}
\fi

\graphicspath{{./}}

\newcommand{\lsim}{\,\lower2truept\hbox{${<\atop\hbox{\raise4truept\hbox{$\sim$}}}$}\,}\newcommand{\gsim}{\,\lower2truept\hbox{${>\atop\hbox{\raise4truept\hbox{$\sim$}}}$}\,}\newcommand{\pp}{~~~.}

\newcommand{\be}{\begin{equation}}\newcommand{\ee}{\end{equation}}\newcommand{\bea}{\begin{eqnarray}}\newcommand{\eea}{\end{eqnarray}}\newcommand{\beann}{\begin{eqnarray*}}\newcommand{\eeann}{\end{eqnarray*}}

\newcommand{\eprint}[1]{\url{arXiv:#1}}

\makeatother

\usepackage{babel}

\makeatother

\usepackage{babel}

\makeatother

\usepackage{babel}

\makeatother

\usepackage{babel}

\makeatother

\usepackage{babel}

\begin{document}

\title[Constraints on coupled dark energy using CMB data from WMAP and SPT]
{Constraints on coupled dark energy using CMB data from WMAP and SPT}


\author{Valeria Pettorino$^{1,3}$, Luca Amendola$^{2}$, Carlo Baccigalupi$^{3,4}$, Claudia Quercellini$^{5}$}
\affiliation{
$^1$ University of Geneva, D\'epartement de Physique Th\'eorique, 24 quai Ernest Ansermet, 1211 Gen\`eve 4, Switzerland.
\\
$^2$ Institut f\"ur Theoretische Physik, Universit\"at Heidelberg,
Philosophenweg 16, D-69120 Heidelberg, Germany.
\\
$^3$ SISSA, Via Bonomea 265, 34136 Trieste, Italy.
\\
$^4$ INFN, Sezione di Trieste, Via Valerio 2, 34127, Trieste, Italy. 
\\
$^5$ University of Rome Tor Vergata, Via della Ricerca Scientifica, 1 - I-00133 Roma, Italy.
}

\begin{abstract}
We consider the case of a coupling in the dark cosmological sector, where a
dark energy scalar field modifies the gravitational attraction between dark matter
particles. We find that the strength of the coupling $\beta$ is constrained using current
Cosmic Microwave Background (CMB) data, including WMAP7 and SPT, to be less than 0.063 (0.11) at $68\%$ ($95\%$)
confidence level. 
Further, we consider the additional effect of the CMB-lensing amplitude, curvature, effective
number of relativistic species and massive neutrinos and show that the bound
from current data on $\beta$ is already strong enough to be rather stable with 
respect to any of these variables. The strongest effect is obtained when we allow 
for massive neutrinos, in which case the bound becomes slightly weaker,  $\beta < 0.084 (0.14)$. 
A larger value of the effective number of relativistic degrees of freedom favors larger
couplings between dark matter and dark energy as well as values of the spectral
index closer to $1$. 
Adding the present constraints on the Hubble constant, as well as from baryon acoustic oscillations 
and supernovae Ia, we find $\beta< 0.050 (0.074)$. In this case we also find an interesting likelihood peak for $\beta=0.041$ (still compatible with 0 at 1$\sigma$).
This peak comes mostly from a slight difference between the Hubble parameter HST result and the WMAP7+SPT best fit.
Finally, we  show that forecasts of Planck+SPT mock data can pin
down the coupling to a precision of better than $1\%$ and  detect whether the marginal peak we find at
small non zero coupling is a real effect.
\end{abstract}

\date{\today}
\maketitle

\section{Introduction}
Cosmic Microwave Background (CMB) probes have recently broadened our knowledge of primordial acoustic oscillations to small angular scales, 
extending previous measurements of temperature power spectrum (Wilkinson Microwave Anisotropy Probe 7, \cite{Komatsu2011}) up to $l \sim 3000$ 
with first compelling evidence of CMB lensing from the South Pole Telescope (SPT, \cite{k11}) and Atacama Cosmology Telescope (ACT, \cite{act_2011}).
The impact of small-scale CMB measurements and gravitational lensing on cosmology 
is relevant \cite{lewis_challinor_2006} and can be used to constrain cosmological parameters and to address one of the major issues of present cosmology, 
 that is to say the nature of dark energy  \citep{verde_spergel_2002, giovi_etal_2003, 2004PhRvD..70b3515A, acquaviva_baccigalupi_2006, hu_etal_2006, Sherwin:2011gv}. 
The simplest framework for dark energy models considers dark energy as a
cosmological constant $\Lambda$, contributing to about $74\%$ of the total
energy density in the universe and providing late time cosmic acceleration, 
while a Cold Dark Matter represents about $21\%$ ($\Lambda$CDM model). Though theoretically in good agreement with present
observations, a cosmological constant is somewhat unpleaseantly
affected by the coincidence and fine-tuning problems which seem unavoidable in
such a framework. Many alternative models have been proposed, though it is fair to say
that so far no one  completely avoids these problems. Some
encouraging arguments have been put forward in the framework of dynamical dark
energy models, where a scalar field (quintessence or cosmon) rolls down a
 suitable potential \cite{wetterich_1988, ratra_peebles_1988} possibly
interacting with dark matter \cite{amendola_2000, pettorino_baccigalupi_2008}
or gravity \cite{Matarrese:2004xa, Perrotta:2002sw} and therefore modifying the growth of structure. Usually, one of the features of
such dynamical dark energy models is to have a non-negligible amount of dark
energy at early times. The amount of early dark energy (\emph{early} referring
to the time of decoupling) influences CMB peaks in various ways and can be
strongly constrained when including small scale measurements, as shown for instance in Refs. \cite{calabrese_etal_2011,
reichardt_etal_2011}. 

In this paper, we consider the case of coupled dark energy models, in which dark
matter particles feel an interaction, additional to gravity, mediated by the
dark energy scalar field. 
Such an interaction introduces effectively a coupling between the evolution of
the dark energy scalar field and dark matter particles. In this sense, this
class of models is both an example in which a non-negligible amount of early
dark energy is present as well as a typical scenario of modified gravity
theories. When seen in the Jordan frame, a coupling between matter and dark
energy can be reformulated in terms of scalar-tensor theories (or $f(R)$  models).
This is exactly true when the contribution of baryons is neglected. In the
Einstein frame, it is common use to neglect a coupling to baryon within coupled
dark energy models, and consider only dark energy - dark matter interactions.
 Alternatively, in the Jordan frame, scalar-tensor theories ($f(R)$ models) require
some sort of screening mechanism (like chameleon \cite{khoury_etal_2004,
Hui:2009kc, 2012arXiv1204.3906U, 2011arXiv1102.5278D} or symmetrons
\cite{2011PhRvD..84j3521H})  that protects the dark energy scalar field and its
mass within high density regions, so that local solar system constraints are
satisfied. 

The strength of the coupling affects CMB  in several ways, changing the
amplitude, the position of the peaks as well as contributing to the Late
Integrated Sachs-Wolfe (ISW) effect (manifest at large length scales) and to
gravitational lensing (appearing at small length scales in the temperature
spectrum). Moreover, the coupling is degenerate with the amount of cold dark
matter $\Omega_{c}$, the spectral index $n$, the Hubble parameter $H(z)$ (see
\cite{amendola_etal_2012} for a review).
After recalling the effects of the coupling on CMB, we use a Monte Carlo
analysis to constrain the coupling combining WMAP and SPT  real data.
Furthermore we extend our analysis to forecasting the constraints that Planck
data are expected to put on the coupling parameter, combined with mock SPT
data.

This paper is organized as follows. In Section II we recall the main features of
coupled dark energy (CDE) cosmologies. In section III we recall effects of the
coupling on the CMB spectrum and describe the methods used, both with regard to
the implementation of the numerical code and the data used for this paper. In
Section IV we derive
the constraints from existing data for several different runs, including effects
of the effective relativistic degree of freedom $N_{eff}$, CMB-lensing, curvature and
massive neutrinos. Here we also forecast the  constraining capability in
presence of the forthcoming Planck data, joined with SPT mock data. Finally, in Section V we derive our
conclusions.

 \section{Coupled Dark Energy} \label{cde}
 
 
 We consider the case in which an interaction is present between dark energy and
dark matter, as illustrated in 
 \cite{Kodama:1985bj, amendola_2000, Amendola:2003wa, amendola_etal_2003,
amendola_quercellini_2003, pettorino_baccigalupi_2008}. Such cosmologies have to
be seen within the framework of modified gravity, since effectively the
gravitational interaction perceived by dark matter particles is modified with respect
to standard General Relativity.
 Coupled dark energy cosmologies considered here are described by the
lagrangian:  
 \be \label{L_phi} {\cal L} =
 -\frac{1}{2}\partial^\mu \phi \partial_\mu \phi - U(\phi) -
 m(\phi)\bar{\psi}\psi + {\cal L}_{\rm kin}[\psi] \,, \ee in which the mass of
 matter fields $\psi$ is a function of the scalar field $\phi
 $. We consider the case in which the DE is only coupled
 to CDM (hereafter denoted with a subscript $c$, while the subscript $b$ will denote baryons). In this
case, the coupling is not affected by tests on the equivalence principle and
solar system constraints and can therefore be stronger than the one with
baryons. 
 The choice
 $m(\phi)$ specifies the coupling and as a consequence the quantity
 $Q_{(\phi) \mu}$ via the expression: \be Q_{(\phi) \mu} = \frac{\partial
   \ln{m(\phi)}}{\partial \phi} \rho_c \, \partial_\mu \phi . \ee 

$Q_{(\phi) \mu} $   acts as a source term in the Bianchi identities: 
     \begin{align}
T_{\nu:\mu}^{\mu} = Q_{(\phi) \mu}  
\label{tensor_conserv_alpha}
\end{align}
   
If no other species is involved in the coupling then $Q_{(c) \mu} = - Q_{(\phi)
\mu}$.
 Various choices of couplings have been investigated in literature, including
constant $\beta $
\cite{Wetterich1994,Amendola1999,Mangano_etal_2003,Amendola2004,Koivisto:2005nr,
Guo:2007zk,Pettorino:2008ez,quercellini_etal_2008,Quartin:2008px,
Valiviita:2009nu} and varying couplings \cite{ Baldi:2010vv}.
 For a constant coupling, typical values of $\beta$ presently allowed by
observations (within current CMB
 data) are within the range $0\le \beta < 0.06$ (at 95\% CL for a constant
coupling and an exponential potential) \cite{amendola_etal_2003,
Amendola:2003wa, amendola_quercellini_2003, Bean:2008ac}, or possibly more
\cite{lavacca2009, Kristiansen:2009yx} if neutrinos are taken into account or
for more realistic time-dependent choices of the coupling.
 Analysis of the
models and constraint on these couplings have been obtained in several other
ways, including spherical collapse (\cite{Wintergerst:2010ui, Mainini:2006zj}
and references therein), higher-order expansions with the time renormalizazion group \cite{Saracco:2009df},
$N$-body simulations \cite{Baldi:2010td,Baldi_etal_2010,Baldi:2010vv} and
effects on supernovae, CMB and cross-correlation of CMB and LSS
\cite{amendola_etal_2003,amendola_quercellini_2003,Bean:2008ac,lavacca2009,
Kristiansen:2009yx,Mainini:2010ng,DeBernardis:2011iw,Xia:2009zzb,martinelli}
together with Fisher matrix forecasts analysis combining power spectrum and Baryonic Acoustic Oscillations 
measurements as expected by the Euclid satellite \footnote{http://www.euclid-ec.org/ and http://sci.esa.int/euclid} and CMB as expected from Planck
\cite{amendola_etal_2012}.
 
 The zero-component of equation (\ref{tensor_conserv_alpha}) gives the
 conservation equations for the energy densities of each species: 
 \bea
 \label{rho_conserv_eq_phi} \rho_{\phi}' &=& -3 {\cal{H}} \rho_{\phi} (1 + w_\phi) - Q_{(\phi)0} \,\,\,\, , \\
 \label{rho_conserv_eq_c} \rho_{c}' &=& -3 {\cal{H}} \rho_{c} + Q_{(\phi)0} \pp
 \nonumber \eea Here we have treated each component as a fluid with
 ${T^\nu}_{(\alpha)\mu} = (\rho_\alpha + p_\alpha) u_\mu u^\nu + p_\alpha
 \delta^\nu_\mu$, where $u_\mu = (-a, 0, 0, 0)$ is the fluid 4-velocity and
 $w_\alpha \equiv p_\alpha/\rho_\alpha$ is the equation of state. Primes denote derivative with respect to conformal time $\tau$.  The class of models considered here corresponds to the choice: \be \label{coupling_const}
 m(\phi) = m_0 e^{-\beta \frac{\phi}{M}} \,, \ee with the coupling term
 equal to \be Q_{(\phi)0} = - \frac{\beta}{M} \rho_c \phi' \,. \ee 
 Equivalently, the scalar field evolves according to the Klein-Gordon equation, which now includes an extra term that depends on CDM energy density:
 \be \label{kg} \phi'' + 2{\cal H} \phi' + a^2 \frac{dV}{d \phi} = a^2 \beta
 \rho_{c} \,\, . \ee
 Throughout this paper we choose an inverse power law potential defined as:
 \be \label{potential}
  V = V_0 \phi^{-\sigma}
 \ee
 with $\sigma$ and $V_0$ constants.

\section{The coupling effect on the CMB power spectrum}
\begin{figure}
\includegraphics[scale=1]{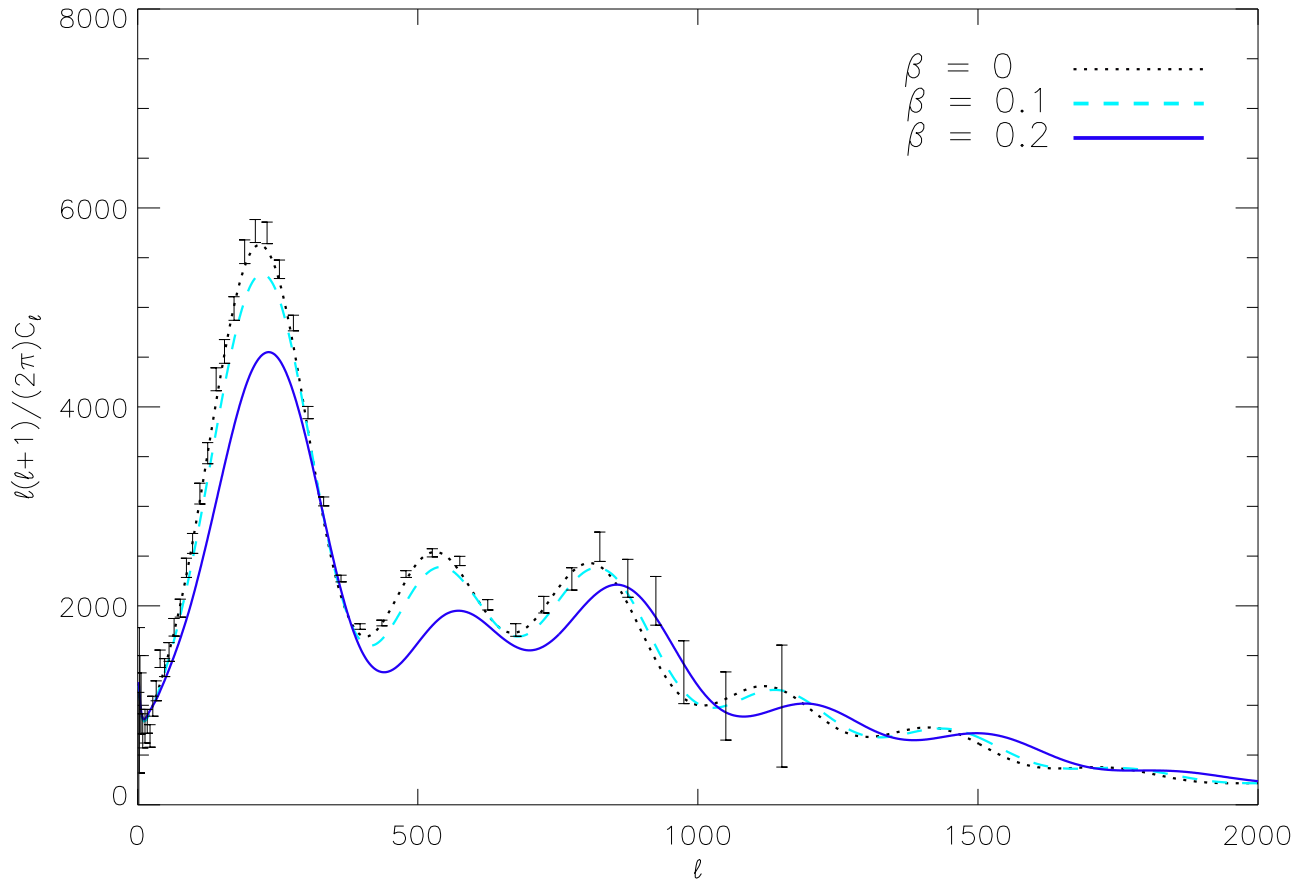}
\caption{\label{fig:CMB-for-three}CMB TT temperature spectra for
three values of $\beta$. Data are taken from WMAP7 \cite{Larson:2010gs}.}
\end{figure}

As discussed in \cite{amendola_etal_2012}, the coupling has two main effects on the CMB: 1) it shifts the position
of the acoustic peaks to larger $\ell$'s due to the increase in the
last scattering surface distance (sometimes called projection effect,
\citep{Pettorino:2008ez} and references therein); 2) it reduces the
ratio of baryons to dark matter at decoupling with respect to its
present value, since coupled dark matter dilute faster than in an
uncoupled model. Both effects are clearly visible in Fig. (\ref{fig:CMB-for-three})
for various values of $\beta$.

%

\begin{figure}
\includegraphics[scale=1]{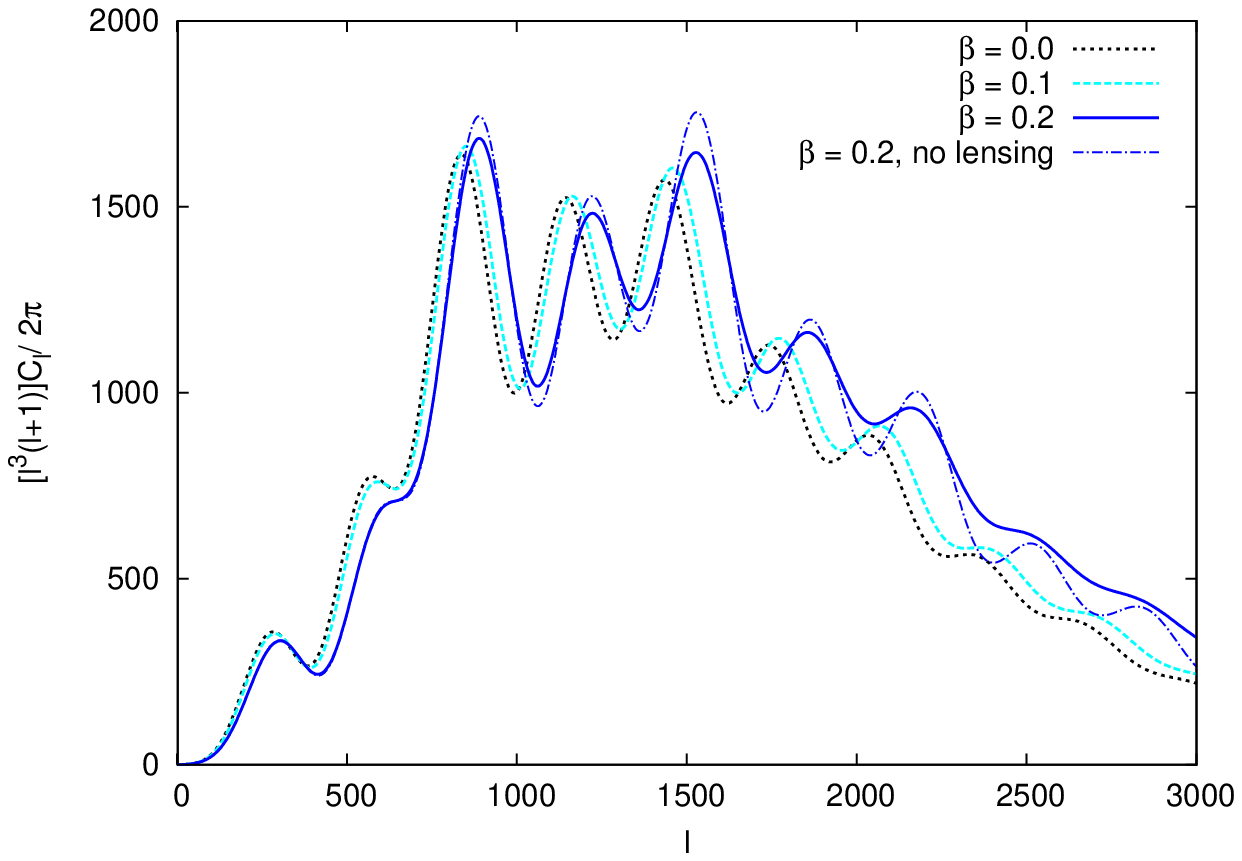}
   \includegraphics[width=13.cm]{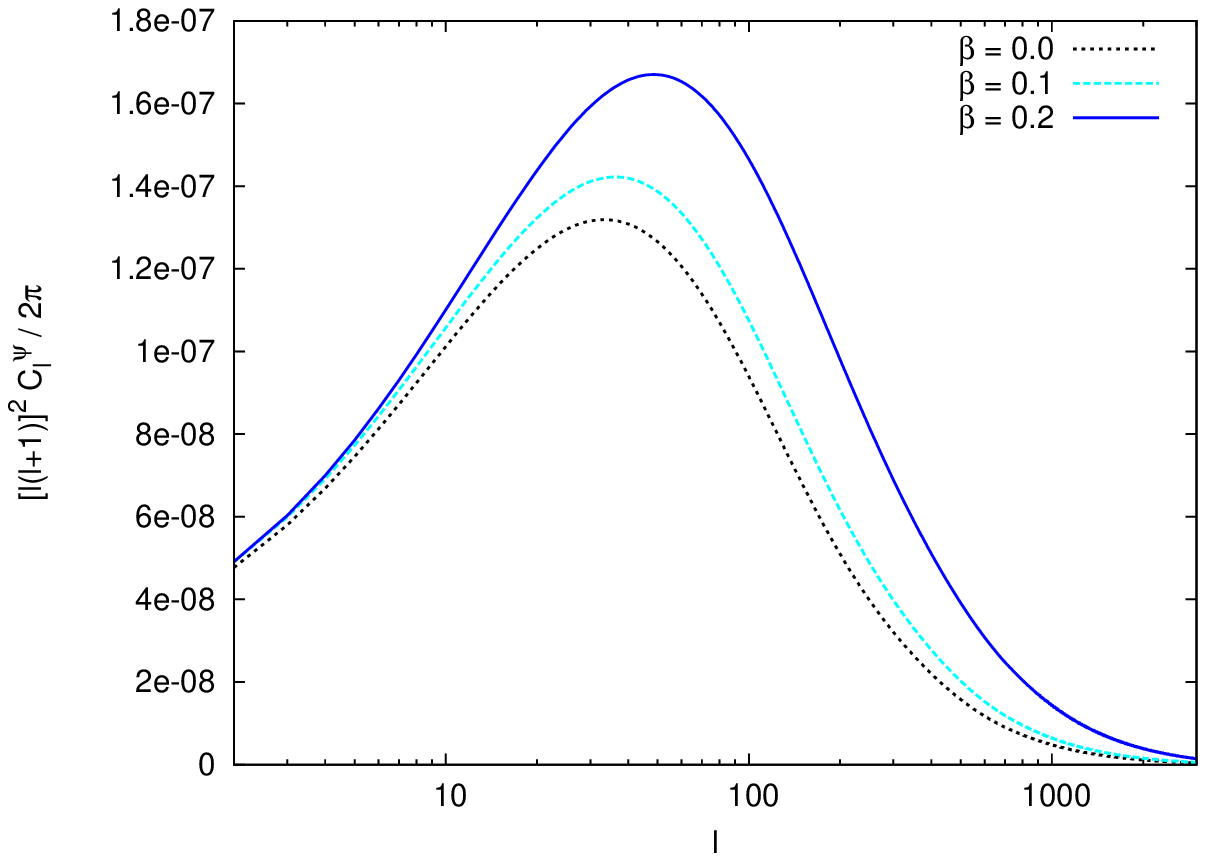}%
\caption{\label{fig:CMB-for-three_timeslsquared_lensing}CMB TT temperature spectra (top panel) and dimensionless lensing potential $l(l+1)C^{\phi \phi}/(2\pi)$  versus multipole $l$ for three values of the coupling $\beta$ .}
\end{figure}

In Fig.\ref{fig:CMB-for-three_timeslsquared_lensing} the effect of the coupling on the CMB is more evident. The figure shows the quantity $C_l(l+1)l^3/(2\pi)$
(i.e. the usual TT spectrum plotted in Fig.(\ref{fig:CMB-for-three}) multiplied by $l^2$), as suggested for example in \cite{Hlozek:2011pc}.
 In Fig.\ref{fig:CMB-for-three_timeslsquared_lensing}  we also show the effect
of including CMB-lensing on the TT spectrum for the same value of the coupling
constant: the unlensed TT spectrum (blue dot dashed) clearly differs from the
corresponding lensed one (solid blue), an effect which is larger at small scales
(large l). The position of the peaks remains invariant but the amplitude is
larger and, for $\ell\lsim 2000$,  the throats appear more pronounced than the peaks.

The purpose of our analysis is to use recent CMB observations reaching multipoles up to $l \sim 3000$ to
constrain dynamical dark energy models in which a coupling is present between
the quintessence scalar field (or cosmon, seen as the mediator of a fifth force
between dark matter particles) and dark matter.
In order to do so, we proceed as follows.

\subsection{Theoretical spectra} 
Theoretical CMB and lensing power spectra have been produced using the code IDEA
(Interacting Dark Energy Anisotropies) based on CAMB \cite{Lewis:1999bs} and
able to include dynamical dark energy, Early Dark Energy parameterizations (not
included in this analysis) as well as  interacting dark energy models.
In order to include the coupling, both background and linear perturbations have
been modified following Refs. \cite{Amendola1999, Pettorino:2008ez}.
The output has been compared to an independent
code \cite{amendola_quercellini_2003} that is built on CMBFAST and
the agreement was better than 1$\%$. The difficulty in the implementation relies
on the fact that the initial conditions cannot be obtained analytically as in
simple dark energy parameterizations (early dark energy or $(w_0, w_a)$):
instead, they must be found by trial and error, through an iterative routine
that finds the initial conditions required to get the desired present values of
the cosmological parameters.

We have then performed a Monte Carlo analysis integrating IDEA within COSMOMC
\cite{cosmomc_lewis_bridle_2002} comparing our theoretical predictions with the
data presented in the next subsection. 

We recall that the CMB coming from the last scattering surface (LSS) is bent by gravitational
structures on the path towards us; this effect is called CMB-lensing 
\citep{Bartelmann:1999yn, lewis_challinor_2006}. 
The standard deviation of the deflection angle is of the order of 2 arcminutes, which would correspond to
small scales and $l > 3000$ multipoles, where CMB peaks are already damped by
photon diffusion. However, deflection angles are correlated with each other over
degree scales, so that lensing can have an important effect on the scales of 
the primary acoustic peaks, mainly smoothing them and transferring power to
larger multipoles. Recently, CMB lensing detection has been claimed by several groups analyzing
total intensity CMB anisotropies \citep{act_2011, k11}. 


CMB lensing is a probe of the expansion rate and naturally depends on the growth of perturbation and on
the gravitational potentials; since dark energy affects both aspects, CMB
lensing represents a way to discriminate among dynamical Dark Energy models and
$\Lambda$CDM, with promising results \citep{verde_spergel_2002, giovi_etal_2003, 2004PhRvD..70b3515A, acquaviva_baccigalupi_2006, hu_etal_2006, Sherwin:2011gv}. 
In particular, the CMB weak lensing theory in generalized cosmologies has been outlined in \cite{2004PhRvD..70b3515A}. 
The difference between the lensed and unlensed curves in Fig.\ref{fig:CMB-for-three} shows the typical effects from CMB lensing. The 
acoustic peaks are smeared because of the correlation between different scales induced by lensing, and for 
the same reason a fraction of power is transferred to the angular domain corresponding to the damping tail, therefore 
dominating that part of the spectrum. Earlier works \cite{acquaviva_baccigalupi_2006} have pointed out how the 
lensing is most relevant in particular in early dark energy models, as it injects power at the onset 
of cosmic acceleration, $z\simeq 1\pm 0.5$ constraining the dark energy abundance in the corresponding epoch. 
On these lines, the analysis in \citep{Sherwin:2011gv} has shown that the inclusion of lensing data
promote the CMB alone to be a probe of the existence of dark energy,
breaking geometrical degeneracies associated to the pure CMB anisotropies at
last scattering.

Also, the lensing depends on time, combining information from decoupling (from
the last scattering surface of the CMB) and $z < 5$ (when large scale structures
formed); recent studies 
\cite{carbone_etal_2009, Teyssier_etal_2009} focus on implementing and investigating simulations of 
CMB lensing through cosmological structures in N-body simulations.

During matter dominated era (MDE), the potentials encountered along the way are constant in the linear regime and the gradient of the potential causes a total deflection angle given by:
\begin{equation}
\alpha = -2 \int_0^{\chi_{*}} d \chi \frac{f_K(\chi_* - \chi)}{f_K(\chi_*)} \nabla_{\perp} \Psi (\chi \hat{n}; \tau_0 - \chi)\ ,
\label{alpha_lensing}
\end{equation} 
where $\chi_*$ is the conformal distance of the source acting as a lens, $\Psi$ is its gravitational potential, $\eta_0 - \chi$ is the conformal time at which the CMB photon was at position $\chi \hat{n}$.

One can also define the gravitational lensing potential
\begin{equation}
\psi(\hat{n}) \equiv -2 \int_0^{\chi_*} d \chi \frac{f_K(\chi_*-\chi)}{f_K(\chi_*)f_K(\chi)} \Psi(\chi \hat{n}; \eta_0 - \chi)\ .
\end{equation}
The lensed CMB temperature $\tilde{T}_{\hat{n}}$ in a direction $\hat{n}$ is
given by the unlensed temperature in the deflected direction $\tilde{T}(\hat{n})
= T(\hat{n}') = T(\hat{n} + \alpha)$ where at lowest order the deflection angle
$\alpha = \nabla \psi$ is just the gradient of the lensing potential.
Expanding the lensing potential into spherical harmonics, one can define also
the angular power spectrum $C_l^\psi$ corresponding to the lensing potential,
defined as $<\psi_{lm} {\psi^*}_{l'm'} > = \delta_{ll'}
\delta_{mm'} C_l^{\psi}$; the latter (multiplied by $[l(l+1)]^2$) is shown in
Fig.\ref{fig:CMB-for-three_timeslsquared_lensing} (lower panel) for different values of the coupling. As we can see
from the plot, the CMB lensing potential mainly gives contribution to large
scales up to $l \sim 1000$ or less. However, the lensed CMB temperature power
spectrum  depends on the convolution between the lensing potential and the
unlensed temperature spectrum (see \cite{lewis_challinor_2006} for more details)
whose effect is of several percent at $l > 1000$, thus being important when
estimating the spectrum up to small scales of $l \sim 3000$, as for the data we consider in the following.

\subsection{Observed data}
We have compared theoretical predictions of CMB lensed spectra with two
datasets. The first includes WMAP7 temperature spectra \cite{Komatsu2011}. The
second includes the recently released power spectrum data from SPT \cite{k11}.
Together with the ATC \cite{act_2011}, SPT
\cite{k11} has recently shown evidence of CMB-lensing with an enhancement in the
CMB temperature power spectrum up to $l \sim 3000$. We do not use ACT data in
this analysis.

\begin{figure*}
\centering
   \includegraphics[width=15.cm]{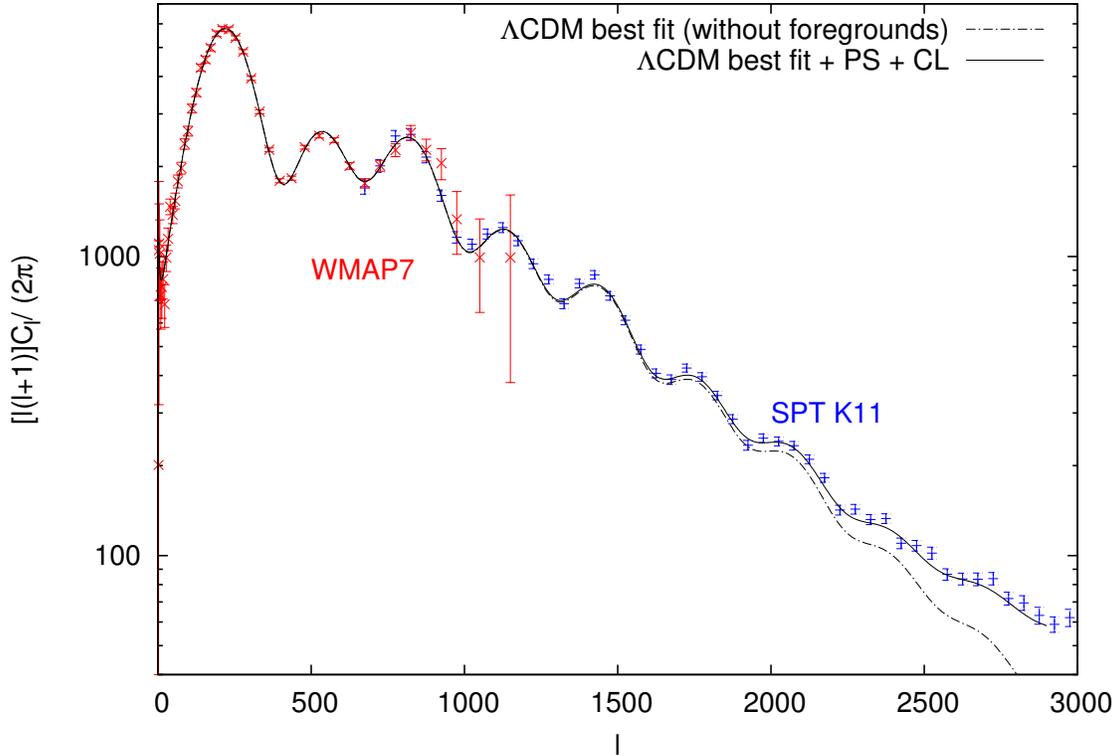}%
     \caption{\small WMAP7 and SPT data used for this work. SZ effects are not included in the best fit line (solid black) (see fig.5 of \cite{k11} for the best fit including SZ). The best fit from WMAP+SPT, for a $\Lambda$CDM without foregrounds is also shown (dash-dotted black line). This plot is similar to fig.5 of \cite{k11} (without SZ in the best fit); we reproduced it here for convenience and because it will allow us to neglect SZ when combining Planck with mock SPT data, in section V. All foregrounds are instead included when SPT is combined with WMAP7.}
\label{fig:cq_data}
\end{figure*}

\begin{figure*}
\centering
   \includegraphics[width=15.cm]{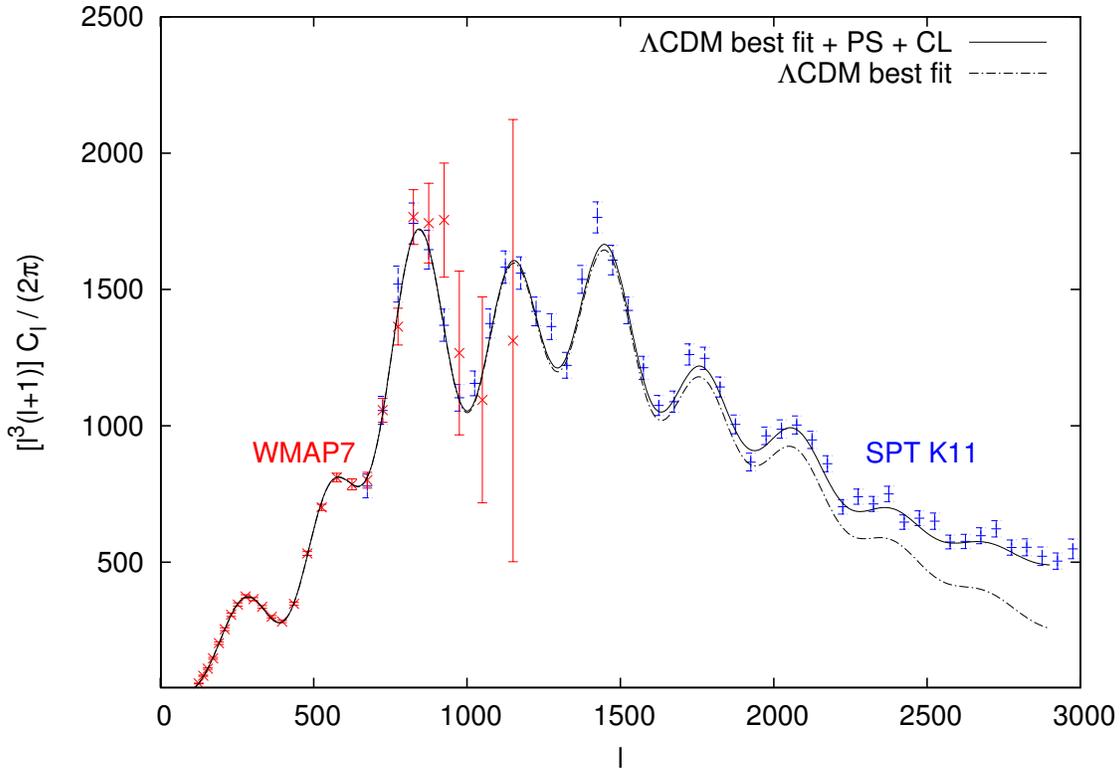}%
     \caption{\small WMAP7 and SPT data used for this work, with an extra factor of $\ell^2$. SZ effects are not included in the best fit line (solid black) (see fig.5 of \cite{k11} for the best fit including SZ). The best fit from WMAP+SPT, for a $\Lambda$CDM without foregrounds is also shown (dash-dotted black line).}
\label{fig:cq_data_l2}
\end{figure*}

In order to use SPT data, we have installed the likelihood provided by the SPT
team \cite{k11} on their SPT website
\footnote{http://pole.uchicago.edu/public/data/keisler11/index.html} and
integrated it with the recommended version of COSMOMC
\footnote{http://cosmologist.info/cosmomc} (August 2011). We have then
implemented IDEA on this version.
Care has to be used whenever small multipoles in the range $2000 - 3000$ are
used, due to several sources of foregrounds active on those scales.
In particular, whenever SPT data are included in the analysis we also
marginalize over the three nuisance parameters described in \cite{k11}: two of
them refer to Poisson point sources and clustered point sources; the third one
adds power from the thermal and kinetic Sunyaev-Zel'dovich (SZ) effects, which
are also an example of secondary CMB anisotropies. 
These effects are relevant when small scales ($l \gsim 2000$) are included and
have to be taken into account whenever SPT data are used.
In Fig.(\ref{fig:cq_data}) and Fig.(\ref{fig:cq_data_l2}) we show the superposition of WMAP7 and SPT data, together with the best fit of the combined set. 

\normalsize

\normalsize

The baseline set of parameters includes $\Theta = {\Omega_b h^2, \Omega_{c} h^2, \Omega_{de},
\theta_s, log{\cal{A}}, n_s, \tau}$. As in other papers, these parameters represent the fractional abundances of the 
various species, as well as the amplitude and shape of the primordial power spectrum, and the reionization optical depth; 
as stressed already, we also consider the three nuisance parameters when SPT is included; when we  impose spatial flatness,  
the present dark energy density  $\Omega_{de}$ becomes a derived parameter; 
in addition, when CQ is included two more parameters are added: $\beta$ and
$\sigma$. Again, $\beta$ represents the coupling between dark matter particles in eq.\ref{coupling_const},
$\sigma$ is the parameter in the scalar field potential  (\ref{potential}) that drives the long
range interaction. Different runs are illustrated in Tab.\ref{tab:runs}. The Helium abundance $Y_{He}$ is derived following BBN consistency (see \cite{k11} for details).
  
%

 \begin{center}
\begin{table}
\begin{tabular}{cccccccc} 
Run & \begin{minipage}{100pt}
\center
CMB-Lensing \\ 
\end{minipage} &  
\begin{minipage}{50pt}
\center
WMAP7
\end{minipage} &
\begin{minipage}{50pt}
\center
SPT
\end{minipage} & 
\begin{minipage}{50pt}
\center
Planck
\end{minipage} & 
\begin{minipage}{50pt}
\center
Parameters
\end{minipage} &
\\
\hline
$ \emph{cq1}  $ & \checkmark & \checkmark & \checkmark & X & baseline + $\sigma$ + $\beta$ \\
$ \emph{cq2}  $ & \checkmark & \checkmark & X & X &baseline + $\sigma$ + $\beta$\\
$ \emph{cq4}  $ & \checkmark & \checkmark & \checkmark &X & baseline + $\sigma$ + $\beta$ + $N_{\nu}$\\
$ \emph{cq1NL}  $ & X & \checkmark & \checkmark &X & baseline + $\sigma$ + $\beta$ \\
$ \emph{cq1K}  $ &  \checkmark & \checkmark & \checkmark &X & baseline + $\sigma$ + $\beta$ + curvature \\
$ \emph{cq1AL}  $ &  \checkmark & \checkmark & \checkmark &X & baseline + $\sigma$ + $\beta$ + $A_l$ \\
$ \emph{cq1}\nu $ &  \checkmark & \checkmark & \checkmark &X & baseline + $\sigma$ + $\beta$ + $f_\nu$ \\
$ \emph{cq1hst} $ &  \checkmark & \checkmark & \checkmark &X & baseline + $\sigma$ + $\beta$ + HST + BAO + SNae\\  
$ \emph{cq1Pl} $ &  \checkmark & X & \checkmark (mock) & \checkmark (mock) & baseline + $\sigma$ + $\beta$ + HST + BAO + SNae\\  
\hline \end{tabular}
\caption{COSMOMC Monte Carlo simulation runs described in this paper; they all refer to CQ models.} \label{tab:runs}
\end{table}
\par\end{center}

\normalsize
 
\section{Results} \label{sec:results}
As a first step, we have  performed a run using WMAP7 and SPT with a
$\Lambda$CDM model, and we find results compatible with \cite{k11}.
Note for the following, that in \cite{k11} the authors report the mean values of
each parameter, together with its standard deviation. We instead report the best
fit values and the marginalized errors at 68$\%$ and 95$\%$ confidence level. We
now describe results from the runs illustrated in Tab.\ref{tab:runs}.

\subsection{Baseline plus $\beta$ and $\sigma$}
The first two Monte Carlo runs we describe (\emph{cq1} and \emph{cq2}) use the baseline
set of parameters $\Theta \equiv \{\Omega_b h^2, \Omega_{c} h^2, \theta_s,
log{\cal{A}}, n_s, \tau \}$; in addition, two more parameters are added to account
for the coupling ($\beta$) and the dark energy scalar field potential
($\sigma$).
Run \emph{cq1} compares theoretical spectra with WMAP7+SPT data; run  \emph{cq2} includes
WMAP7 data only. Results are shown in Tab.\ref{tab:cq1cq2_bestfit}, where we report best fit values with 1-sigma (68\%) errors on various parameters. For $\beta$ we also
report upper 1 and 2 $\sigma$ limits.
When WMAP7 only is considered, the coupling is constrained to be $\beta <
0.078 (0.14)$ at 1 (2) $\sigma$, while including SPT data, and therefore small scales and large
multipoles, the coupling is constrained down to $< 0.063 (0.11)$. 

In the past, \cite{amendola_etal_2003} found $\beta_A < 0.16 $ at 95\% c.l.
(note that our definition of $\beta$ is $\beta = \sqrt(2/3) \beta_A = 0.13$ at 2
$\sigma$) using 
COBE, Boomerang, Maxima, DASI and fixing the optical depth $\tau$.
\cite{amendola_quercellini_2003} found $\beta < 0.061 (0.11)$ at 1 (2) $\sigma$
using WMAP1 data and $\beta < 0.11 (0.16)$ for pre-WMAP data (using a different
set than \cite{amendola_etal_2003}).
Ref. \cite{Bean:2008ac} found C < 0.034 (0.066) (in their notation $C \equiv
\beta_A \equiv \sqrt{2/3} \beta$ so that their result is equivalent to $\beta <
0.028 (0.054)$) using WMAP5, SNLS, HST, LRG, SDSS.
See also \cite{Xia:2009zzb} (with a different definition of the coupling) and
\cite{lavacca2009} (where massive neutrinos were included).
These constraints are not easily comparable to ours since in these papers different
priors have been used and/or some parameters have been kept fixed.
Overall, however, we find an agreement with the more recent  constraints to within a factor
of 50\% at most.

\begin{figure*}
 \centering
 \includegraphics[width=18.cm]{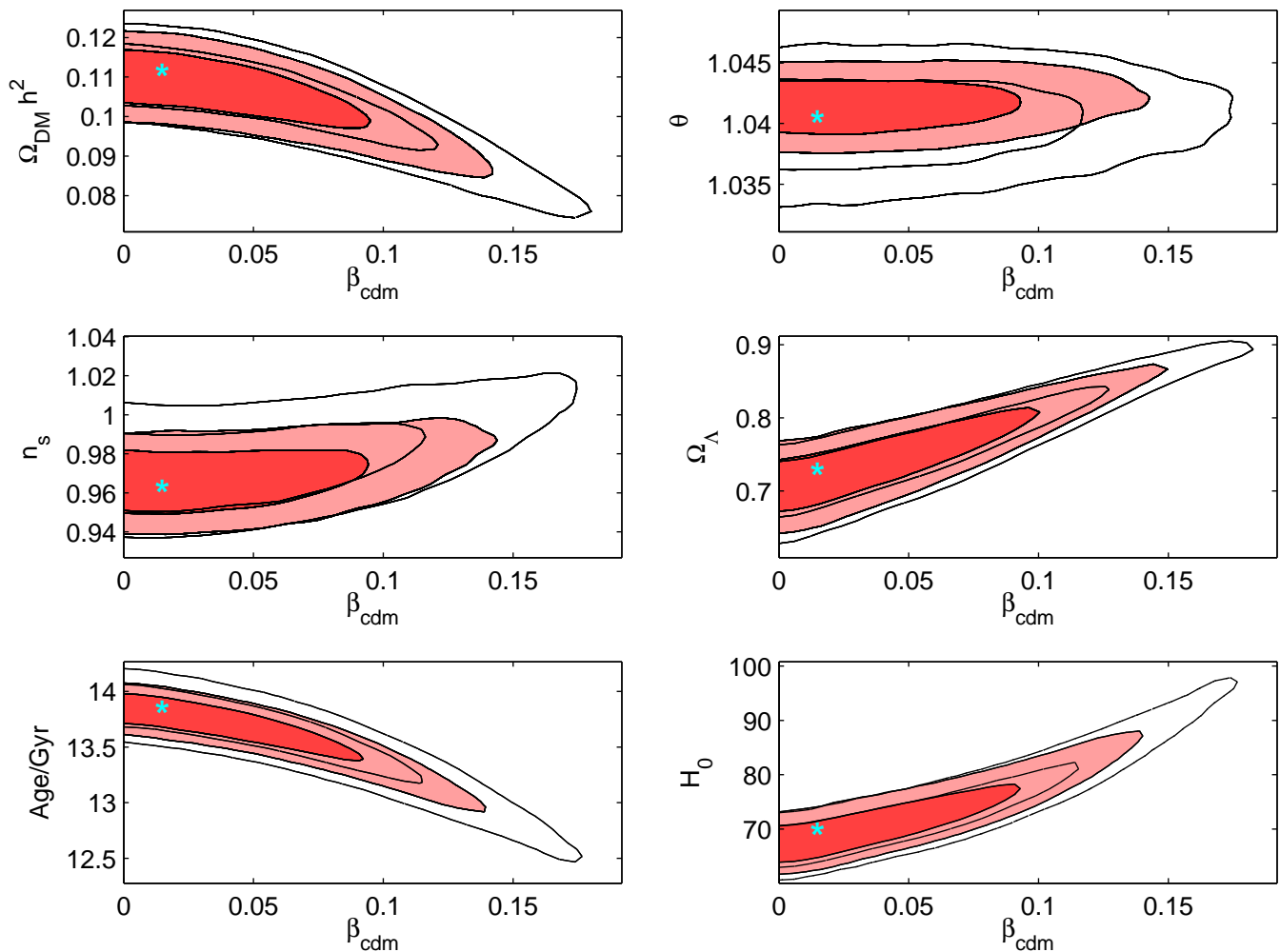}
      \caption{\small Confidence contours for the cosmological parameters for coupled quintessence models. We compare runs \emph{cq1} (red) and \emph{cq2} (white). The light blue asterisks mark the best fit points for $cq1$.  1-sigma and 2-sigma contours are shown.}
 \label{fig:like_cont_baseline_cq1_cq2}
 \end{figure*}

\begin{figure*}
 \centering
    \includegraphics[width=17.cm]{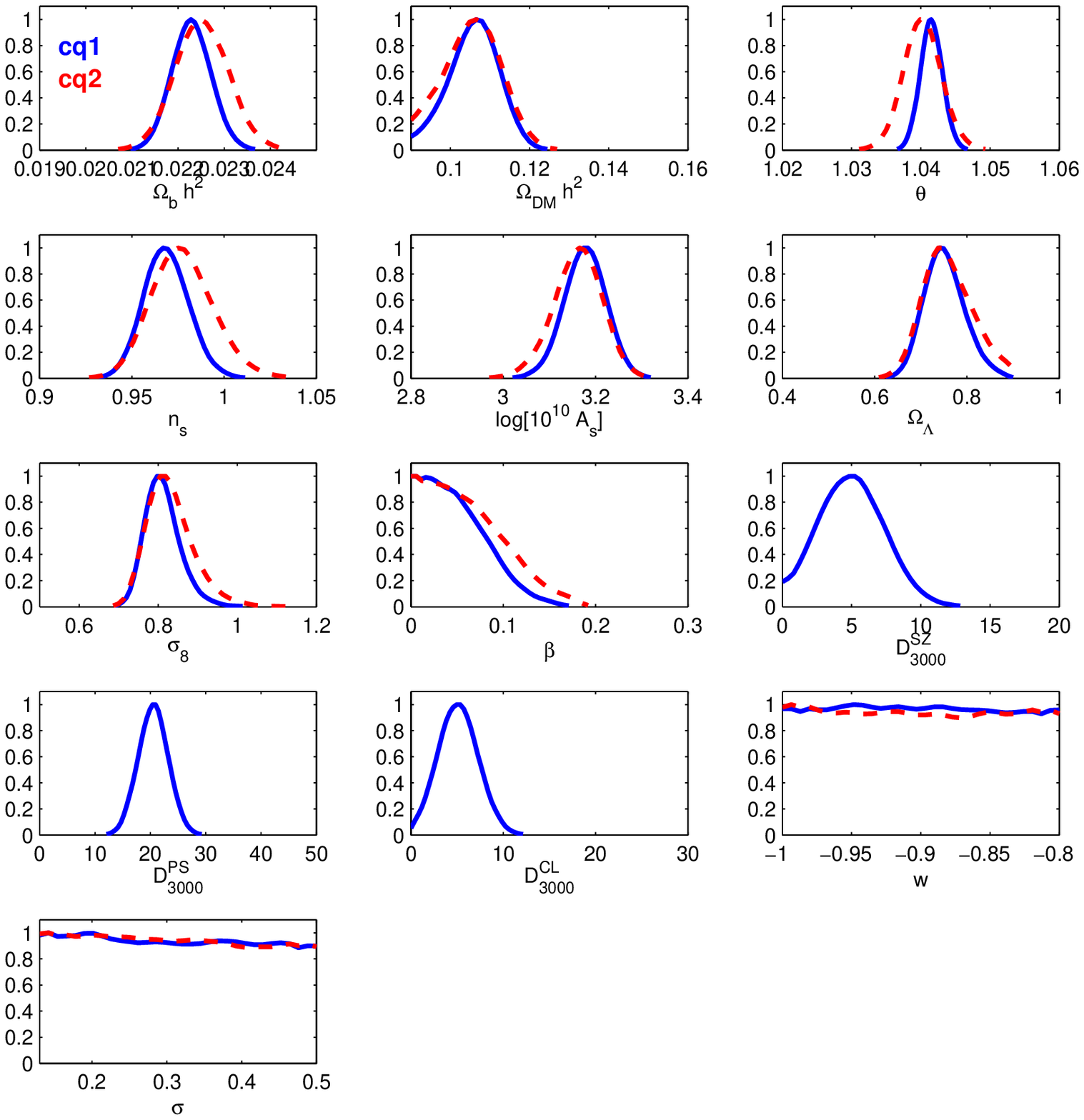}%
      \caption{\small 1D likelihoods for the cosmological parameters for coupled quintessence models. We compare runs  \emph{cq1} (solid blue) and  \emph{cq2} (dashed red). As expected, in coupled quintessence models, no dependence is seen on the value of $\sigma$ and $w$, which are arbitrary. We restrict the analysis to a reasonable range in $w$ and a range in $\sigma$ small enough to speed up the iterative routine that finds initial conditions.}
 \label{fig:like_cont_1D_cq1_cq2}
 \end{figure*}

The 2D confidence contours are plotted in Fig.\ref{fig:like_cont_baseline_cq1_cq2}.
Here we show a selection of the most interesting likelihood contours vs the
coupling $\beta$.
In Fig.\ref{fig:like_cont_1D_cq1_cq2} we also show 1D likelihood contours,
comparing results from WMAP7+ SPT with WMAP7 only. Note that there is no
dependence of cosmological parameters from $\sigma$, as expected since $\sigma$ only affects late time cosmology; the
range in $\sigma$ was therefore safely chosen to be between 0.13 and 0.5, small enough to get
reasonable speed for the runs. The value of $w$ is arbitrary and approximately related to
$\sigma$ via the expression: $w = - 2 / (\sigma + 2)$; the interval chosen for
$\sigma$ is such that $w$ still assumes reasonable values, at least smaller than
-0.8. As it appears clearly from Fig. \ref{fig:like_cont_1D_cq1_cq2}, CMB is practically insensitive
to $\sigma$ or $w$ within the range we consider.
 \begin{center}
\begin{table}
\begin{tabular}{lllll}
\textbf{Best fit values} & \textbf{for coupled quintessence} & \\
\hline
\begin{minipage}{30pt}
\flushleft
Parameter 
\\
\end{minipage} & 
\begin{minipage}{150pt}
\flushleft
cq1 (WMAP7 + SPT)
\end{minipage} &  
\begin{minipage}{150pt}
\flushleft
cq2 (WMAP7)
\end{minipage} &
\\
\\
\hline
$  \Omega_b h^2 $ & $0.022^{ +0.0007}_{-0.00013}$           & $0.023^{+0.00044}_{-0.00070}$		\\
$ \Omega_{c} h^2  $ &$0.11^{+0.0022}_{-0.011}$                                  & $0.11^{+0.0019}_{-0.016}$ \\
      $ \theta_s$ & $1.04^{+0.0024}_{-0.00072}$                    & $1.04^{+0.0027}_{0.0025} $ \\
   $ \tau$ & $0.091^{+0.0013}_{-0.012} $                             & $0.089^{0.0076}_{0.0072}$\\   
   $ n_s$ & $0.96^{+0.019}_{-0.0056} $                                                     & $0.97^{+0.021}_{-0.012}$ \\
   $w$ &  $-0.88^{+0.080}_{-0.12}  $                                                                                 & $-0.97^{+0.17}_{-0.03}$ \\  
   $\beta$ &  $0.012^{+0.050}_{-0.012}$                          & $0.0066^{0.071}_{-0.0066}$\\  
      $\beta$ &  $< 0.063 (0.11) $                          & $< 0.078 (0.14)$\\  
    $ \sigma$ & $0.22^{+0.28}_{-0.090}$                                              & $0.13^{+0.37}_{-0.0048}$ \\
   $ \Omega_{de}$ & $0.72^{+0.076}_{-0.012}$                                  & $0.72^{+0.093}_{-0.016} $ \\
   $ Age/Gyr$ & $13.8^{+0.012}_{-0.39} $                                              & $13.8^{+0.07}_{-0.5}$ \\
   $ z_{re}$ & $10.9^{+0.63}_{-1.8} $                          & $10.6^{+1.2}_{-1.2}$ \\
   $ H_0$ & $68.7^{+8.4}_{-0.98}$                             & $ 69.3^{+10.7}_{-1.7} $  \\ 
            $ D_{3000}^{SZ}$ & $4.0^{+2.1}_{-4.0}$                    & - \\
                  $ D_{3000}^{PS}$ & $21.5^{+1.7}_{-3.7}$                    &- \\
                           $ D_{3000}^{CL}$ & $5.3^{+1.9}_{-2.4} $                    & - \\
   -$Log$(Like) &$ 3756$                      							& $3737 $  \\ 
\hline
\end{tabular}
\caption{Best fit values and 1-$\sigma$ errors comparing runs \emph{cq1} and \emph{cq2}. Both runs include coupling; cq1 uses WMAP7 + SPT while \emph{cq2} uses WMAP7 only. For $\beta$ we also write in brackets the value of the 2$\sigma$ marginalized error.}
 \label{tab:cq1cq2_bestfit}
\end{table}
\par\end{center}
\normalsize

 \normalsize

 \normalsize

\subsection{Effective number of relativistic species $N_{eff}$}
\begin{figure*}
 \centering
    \includegraphics[width=18.cm]{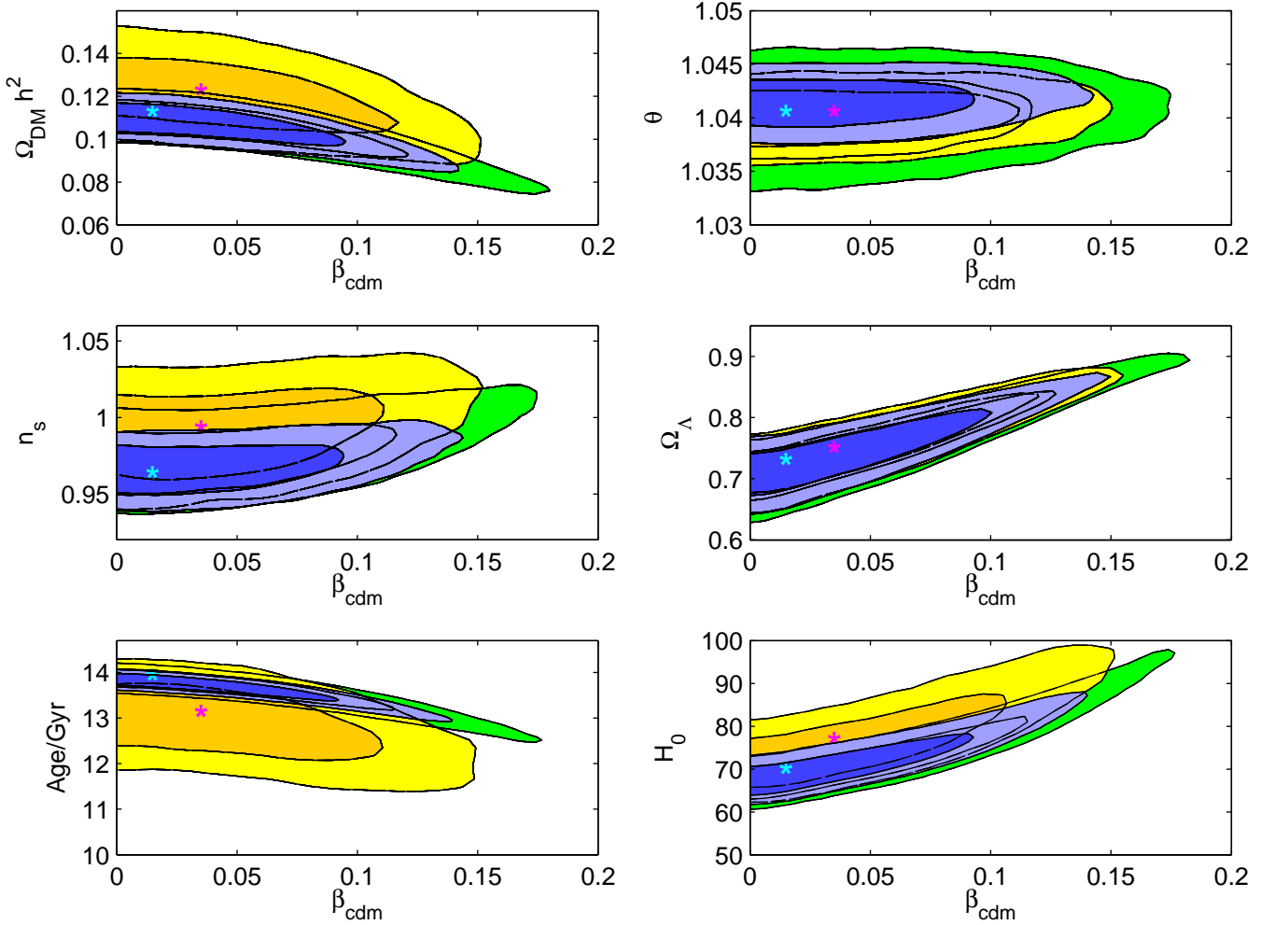}%
      \caption{\small Confidence contours for the cosmological parameters for coupled quintessence models. We compare runs  \emph{cq1} (blue),  \emph{cq2} (green) and \emph{cq4} (yellow). The light blue asterisks mark the best fit points for $cq1$ while the pink asterisks mark the best fit points for $cq4$. 1-sigma and 2-sigma contours are shown.}
 \label{fig:like_cont_neff}
 \end{figure*}

The effective number of relativistic species before recombination, usually
denoted by $N_{eff}$
is higher than the number of relativistic neutrino species (3.046) due to
photons produced in electron-positron annihilation at the end of neutrino
freeze-out \cite{dicus_etal_1982,lopez_etal_1999, mangano_etal_2005,
2011JCAP...03..035M}.
Using WMAP7 + SPT data for a $\Lambda$CDM, $N_{eff} = 3.046$ was found to be
preferred over zero relativistic species ($N_{eff} = 0$) \cite{k11}. 
If $N_{eff}$ is left free to vary and marginalized over, its best fit value can
be even larger: \cite{Komatsu2011} found $N_{eff} > 2.7$ at $95\%$ CL using
WMAP7 alone; \cite{dunkley_etal_2010}  found $N_{eff} = 5.3 \pm 1.3$ using WMAP7
+ ACT; \cite{k11} found $N_{eff} = 3.85 \pm 0.62$ using WMAP7 + SPT. 
If relativistic species are present, the expansion rate
during radiation dominated era increases \cite{hu_white_1996, 
Bashinsky_Seljak_2004, 2011arXiv1104.2333H,  k11}.
We have redone the analysis in the case of coupled quintessence, to check whether the number of relativistic degrees of freedom is degenerate with the coupling. 
The effect of marginalizing over $N_{eff}$ on the coupling is shown in
Tab.\ref{tab:cq1cq4_bestfit}. In Fig.\ref{fig:like_cont_neff} we plot the
likelihood contours for a selection of parameters, vs the coupling $\beta$,
comparing different runs.

\begin{figure*}
 \centering
    \includegraphics[width=18.cm]{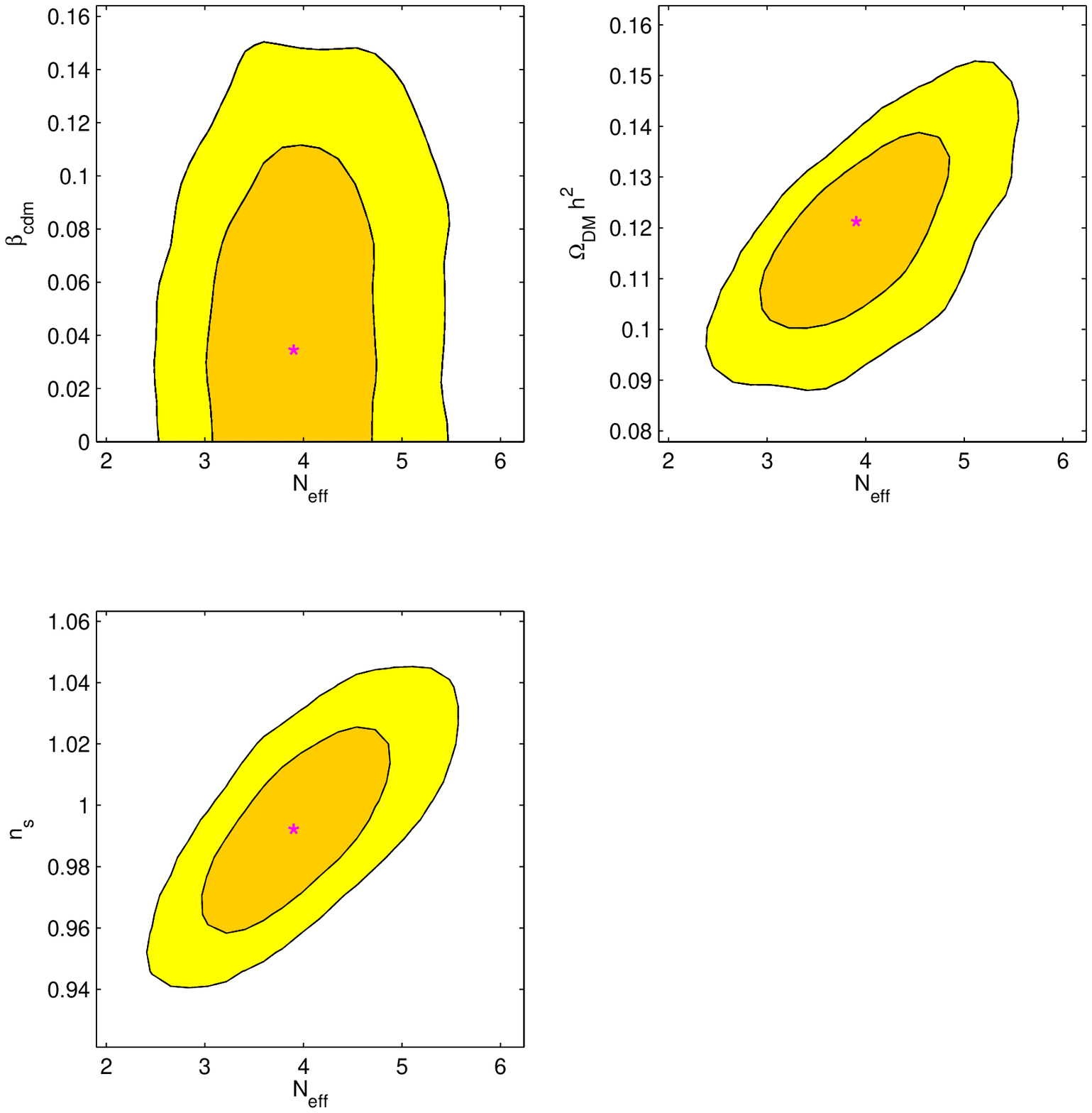}%
      \caption{\small Confidence contours for a choice of cosmological parameters vs $N_{eff}$ within coupled quintessence models for run  \emph{cq4}. 1-sigma and 2-sigma contours are shown..  $N_{eff}$ is degenerate with $\Omega_{DM}h^2$, $n_s$, which are in turn degenerate with $\beta$. No direct degeneracy can however be seen between $\beta$ and $N_{eff}$. The pink asterisks mark the best fit for $cq4$.}
 \label{fig:like_cont_neff_yhe}
 \end{figure*}

Contours are larger when we allow $N_{eff}$ to vary, but the range in $\beta$ is
not affected considerably ($\beta < 0.074 (0.12)$ instead of $\beta < 0.063 (0.11)$). When a
coupling is present, we find that the best fit for the number of relativistic
species is given by $N_{eff} = 3.84^{+0.74}_{-0.49}$ when using WMAP and SPT, similar
to the value mentioned before and evaluated in absence of a coupling.  The
allowed range for $Y_{He}$, obtained through BBN consistency, increases a lot when $N_{eff}$ is free to vary 
\cite{Bashinsky_Seljak_2004}. $N_{eff}$ is degenerate with dark matter
and the spectral index, which in turn are weakly degenerate with $\beta$, though no
direct degeneracy appears between $\beta$ and $N_{eff}$, as shown in Fig.\ref{fig:like_cont_neff_yhe}.

It is interesting to see (Fig.\ref{fig:like_cont_neff}) that when we allow for an effective number of relativistic degrees of freedom, marginalizing over $N_{eff}$, the coupling from WMAP7+SPT data increases to a best fit value of $\beta \sim 0.03$. A larger value of $N_{eff}$ (best fit $\sim 3.8)$ favors larger couplings between dark matter and dark energy as well as values of the spectral index closer to -1 ($n_s \sim 0.99$).

\begin{center}
\begin{table}
\begin{tabular}{lllll}
\textbf{Best fit values} & \textbf{for coupled quintessence} & WMAP7 + SPT\\
\hline
\begin{minipage}{30pt}
\flushleft
Parameter 
\\
\end{minipage} & 
\begin{minipage}{150pt}
\flushleft
 \emph{cq1} (baseline + $\beta$+$\sigma$)
\end{minipage} &  
\begin{minipage}{150pt}
\flushleft
 \emph{cq4} (baseline + $\beta$+$\sigma$+$N_{eff}$)
\end{minipage} &
\\
\\
\hline   
$  \Omega_b h^2 $ & $0.022^{ +0.0007}_{-0.00013}$           & $0.023^{+0.00054}_{-0.00054}$		\\
$ \Omega_{c} h^2  $ &$0.11^{+0.0022}_{-0.011}$                                  & $0.12^{+0.0076}_{-0.016}$ \\
      $ \theta_s$ & $1.04^{+0.0024}_{-0.00072}$                    & $1.04^{+0.0012}_{-0.0025} $ \\
   $ \tau$ & $0.091^{+0.0013}_{-0.012} $                             & $0.086^{+0.012}_{-0.0030}$\\   
   $ n_s$ & $0.96^{+0.019}_{-0.0056} $                                                     & $0.99^{+0.029}_{-0.014}$ \\
   $w$ &  $-0.88^{+0.080}_{-0.12}  $                                                                                 & $-0.89^{+0.090}_{-0.11}$ \\  
   $\beta$ &  $0.012^{+0.050}_{-0.012}$                          & $0.032^{+0.042}_{-0.032}$\\  
      $\beta$ &  $< 0.063 (0.11) $                          & $< 0.074 (0.12)$\\  
    $ \sigma$ & $0.22^{+0.28}_{-0.090}$                                              & $0.14^{+0.36}_{-0.11}$ \\
   $ N_{eff}$ & -  & 						$3.84^{+0.74}_{-0.49}$  \\
   $ \Omega_{de}$ & $0.72^{+0.076}_{-0.012}$                                  & $0.74^{+0.060}_{-0.031} $ \\
   $ Age/Gyr$ & $13.8^{+0.012}_{-0.39} $                                              & $13.0^{+0.40}_{-0.77}$ \\
   $ z_{re}$ & $10.9^{+0.63}_{-1.8} $                          & $10.7^{+1.7}_{-1.0}$ \\
   $ H_0$ & $68.7^{+8.4}_{-0.98}$                             & $ 75.6^{+9.9}_{-4.3} $  \\ 
            $ D_{3000}^{SZ}$ & $4.0^{+2.1}_{-4.0}$                    & $6.7^{+2.2}_{-3.3}$ \\
                  $ D_{3000}^{PS}$ & $21.5^{+1.7}_{-3.7}$                    & $20.2^{+3.0}_{-2.5}$  \\
                           $ D_{3000}^{CL}$ & $5.3^{+1.9}_{-2.4} $                    &   $4.4^{+3.2}_{-1.2} $   \\
   -$Log$(Like) &$ 3756$                      							& $3756 $  \\    
\hline
\hline
\end{tabular}
\caption{Best fit values and 1-$\sigma$ errors comparing runs  \emph{cq1} and  \emph{cq4}. Both runs include coupling and use WMAP7 + SPT; in addition,  \emph{cq4} marginalizes over $N_{eff}$. For $\beta$ we also write in brackets the value of the 2$\sigma$ marginalized error.}
 \label{tab:cq1cq4_bestfit}
\end{table}
\par\end{center}
\normalsize

 \normalsize


 \normalsize

\subsection{CMB lensing}

\begin{figure*}
 \centering
    \includegraphics[width=18.cm]{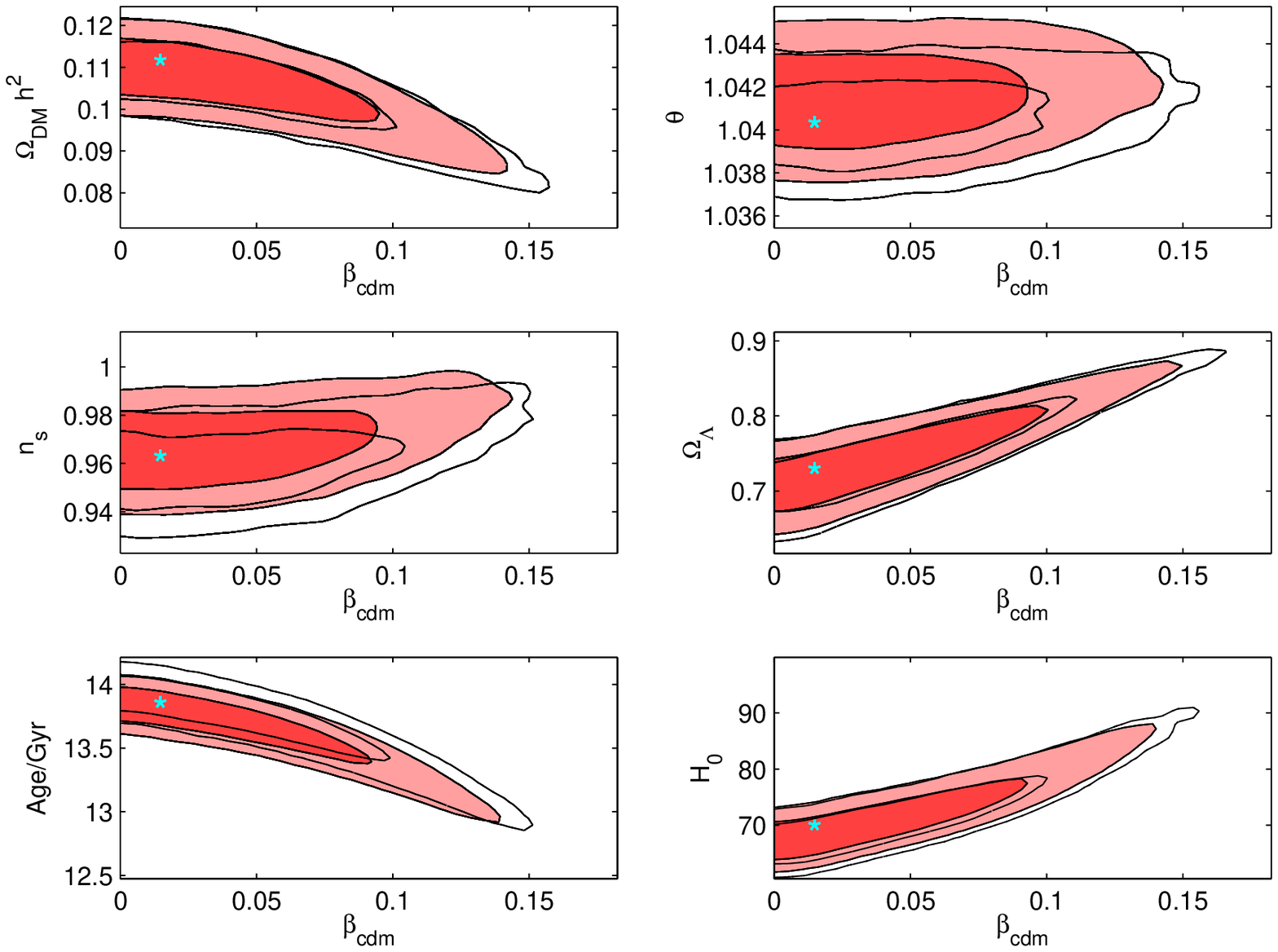}%
      \caption{\small Likelihood contours for cosmological parameters in presence of a coupling for WMAP+SPT (run  \emph{cq1}, including lensing, red contours), as compared to the same run done without CMB lensing (white contours). Light blue asterisks mark the best fit points for $cq1$.}
 \label{fig:like_cont_NL}
 \end{figure*}

In order to test the effect of CMB-lensing on coupled quintessence, we have
redone a run  \emph{cq1} (WMAP + SPT) without lensing.
The presence of a constant coupling doesn't seem to be very much affected by
lensing in the TT CMB spectra, as we can see in Fig.(\ref{fig:like_cont_NL})
where we compare run  \emph{cq1} with run  \emph{cq1NL}. If no lensing is included, the bound on
$\beta$ is slightly (but not significantly) larger: $\beta < 0.068 (0.13)$. 

\begin{figure*}
 \centering
    \includegraphics[width=15.cm]{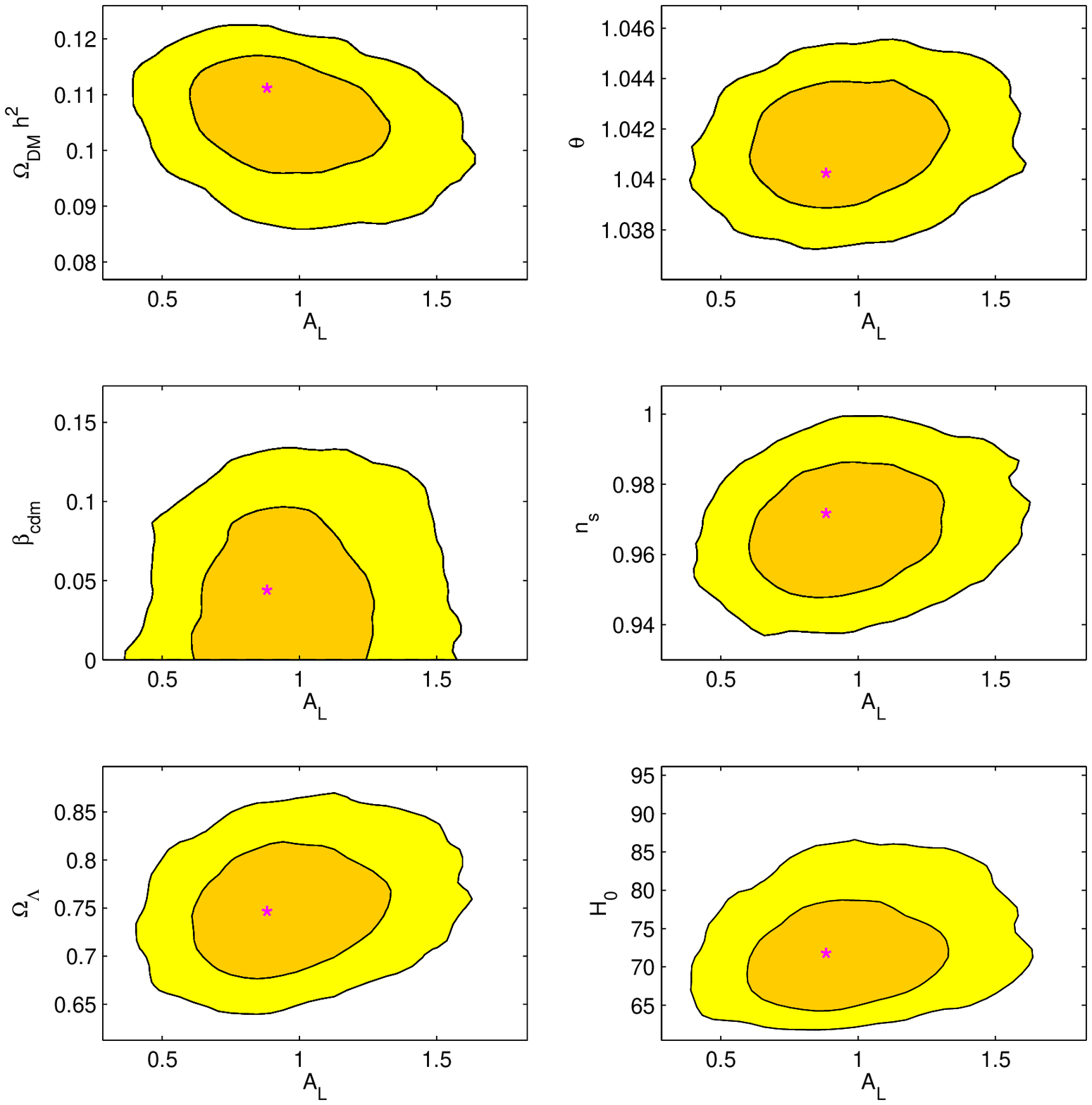}
      \caption{\small Likelihood contours for cosmological parameters when $A_L$ is allowed to vary.}
 \label{fig:like_cont_al}
 \end{figure*}

Furthermore, similarly to \cite{k11} one can rescale the lensing potential power
spectrum by a factor $A_L$:
\begin{equation}
C_{l}^{\phi\phi} \rightarrow A_L C_l^{\phi \phi}
\end{equation}
All runs discussed so far fix $A_L = 1$. In order to test the effect of lensing
we also performed a run ($cq1AL$) in which we vary $A_L$ and marginalize over it. The
$A_L$ parameter was found to be $A_L = 0.94 \pm 0.15$ when using WMAP + SPT with
a $\Lambda$CDM model \cite{reichardt_etal_2011} (see also \cite{2009ApJ...694.1200R,
calabrese_etal_2008, das_etal_2011a, act_2011} for different datasets). When a
coupling between dark matter and dark energy is included, we find that the best
fit for $A_L$ is $A_L = 0.86^{+0.34}_{-0.12}$, still compatible with one. The bound on
the coupling is of the same order as in the case in which $A_L$
is fixed: $\beta < 0.063 (0.11)$. In other words, we don't gain much marginalizing over
$A_L$ instead of fixing it to one, given that $A_L$ best fit is very close and
fully compatible with one. Though with the data considered here the effect is
not significant, $A_L$ is also correlated with dark matter and $n_s$, which in
turn are correlated with $\beta$, as shown in fig.\ref{fig:like_cont_al}.

 \normalsize

 \normalsize

\subsection{Curvature}
\begin{figure*}
 \centering
    \includegraphics[width=18.cm]{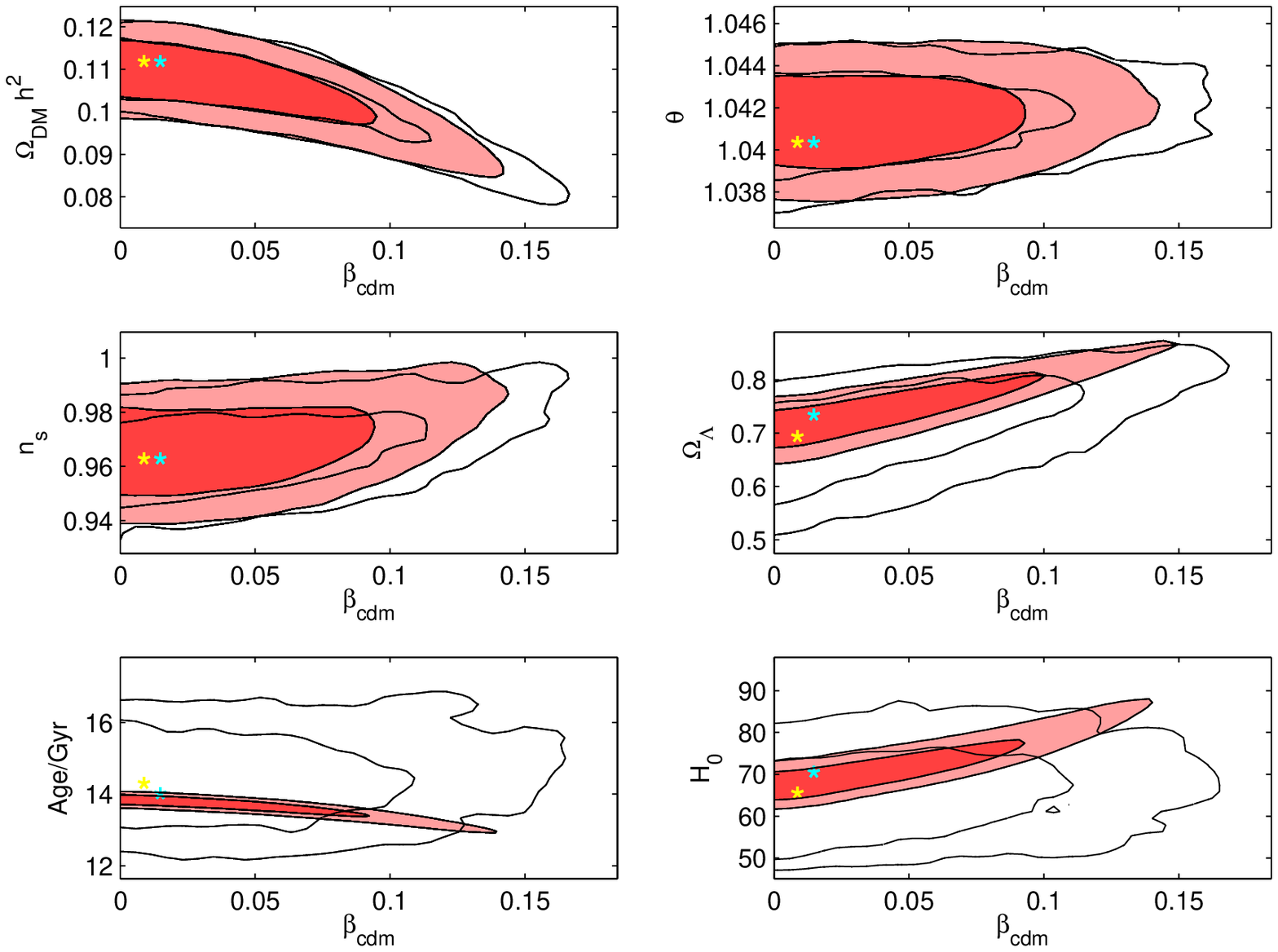}
      \caption{\small Likelihood contours for cosmological parameters of run  \emph{cq1} (red contours) as compared to run  \emph{cq1}, in which $\Omega_K$ is allowed to vary (white contours). Light blue asterisks mark the best fit points of $cq1$ while yellow asterisks mark the best fit points for $cq1K$.}
 \label{fig:like_cont_curvature}
 \end{figure*}

\begin{figure*}
 \centering
    \includegraphics[width=15.cm]{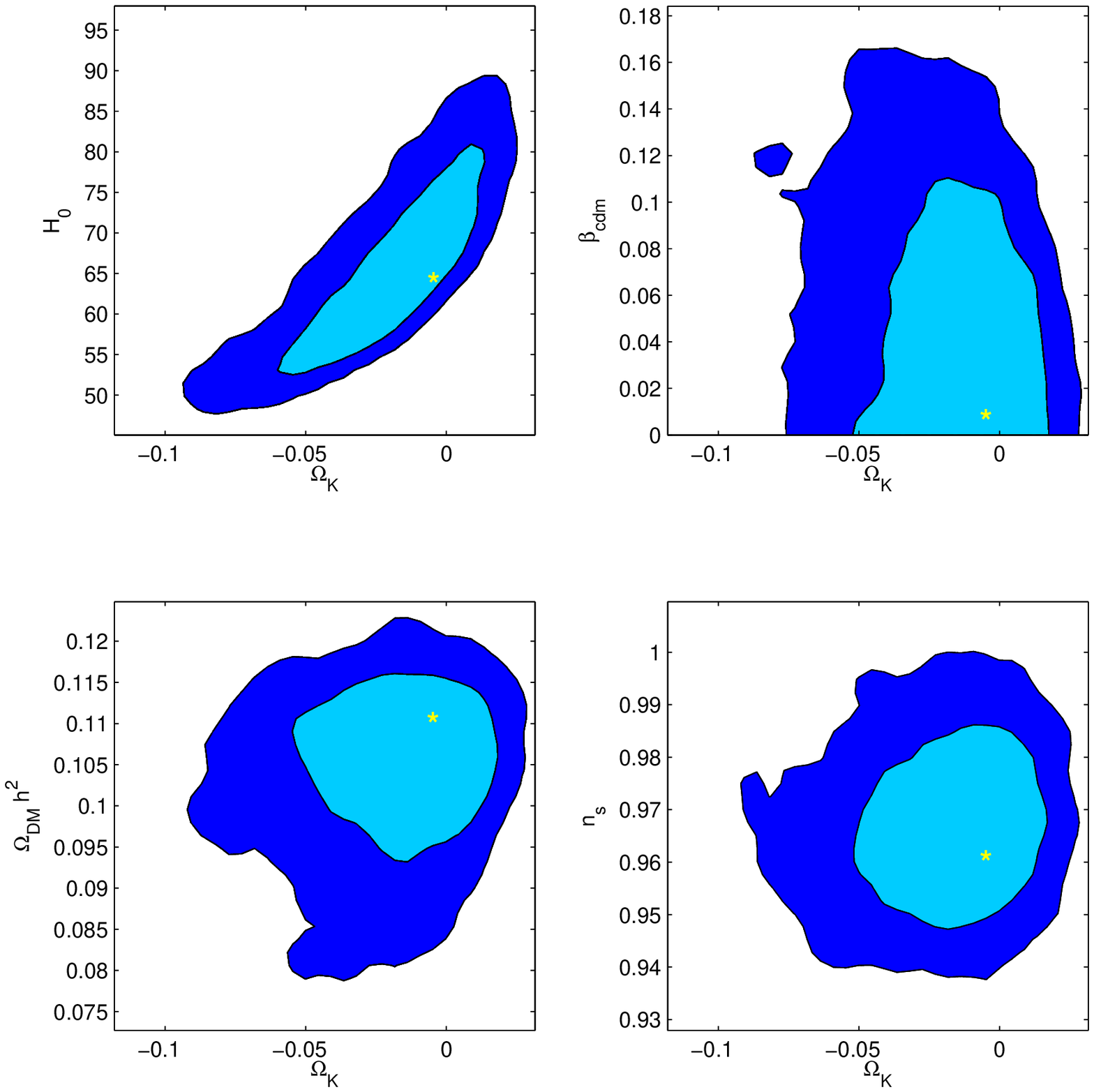}
      \caption{\small Likelihood contours for cosmological parameters when $\Omega_K$ is allowed to vary. Yellow asterisks mark best fit points for run $cq1K$.}
 \label{fig:like_cont_omegaK}
 \end{figure*}

If we release the constraint of a flat universe and allow for curvature and
coupling (run $cq1K$), we get $\Omega_K = -0.0068^{+0.0092}_{-0.036}$, which is compatible with a flat
Universe. In this case, the constraint on $\beta$ is slightly less restrictive,
$\beta < 0.071 (0.13)$ but the bound on $\beta$ is already stringent enough not
to be affected so much by the uncertainty on curvature. Contours are shown in
Fig.\ref{fig:like_cont_curvature} where they are compared to run  \emph{cq1}, in which a flat
universe was assumed. We also show in Fig.\ref{fig:like_cont_omegaK} how
curvature is degenerate, as expected, with the Hubble parameter, though no
direct degeneracy is seen between $\Omega_K$ and the coupling $\beta$.

 \normalsize

 \normalsize

\subsection{Massive neutrinos}
\begin{figure*}
 \centering
    \includegraphics[width=15.cm]{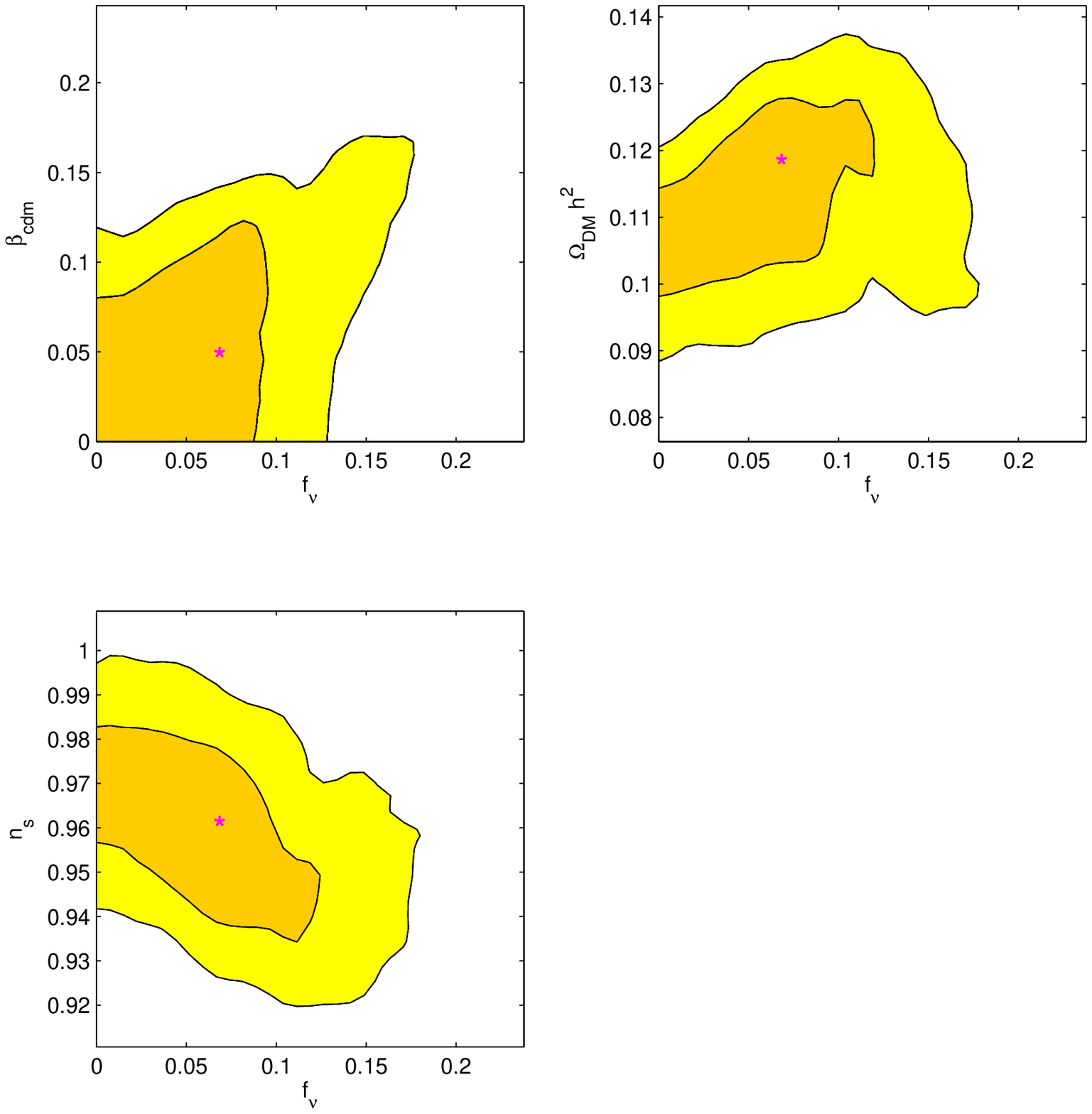}
      \caption{\small Likelihood contours for cosmological parameters when $f_\nu$ is allowed to vary. Pink asterisks mark best fit points for run $cq1\nu$.}
 \label{fig:like_cont_fnu}
 \end{figure*}

Up to now, we have fixed the relative fraction of massive neutrinos $f_\nu$ to
zero. We now consider run  \emph{cq1}$\nu$ in which we also allow for a non zero
fraction of massive neutrinos and marginalize over $f_\nu$. In this case the
range allowed for the coupling is $\beta < 0.084 (0.14)$, slightly bigger than when
using massless neutrinos, as expected \cite{lavacca2009}. We update the results of \cite{lavacca2009}
 using both WMAP7 and SPT data. The degeneracy between massive neutrinos
and $\beta$ is clearly shown in Fig.\ref{fig:like_cont_fnu}. The best fit value for $f_\nu$ is $f_\nu = 0.065^{+0.017}_{-0.065}$

 \normalsize

\subsection{Combining WMAP7 and SPT with HST, BAO and Supernovae Ia data}

\begin{figure*}
 \centering
    \includegraphics[width=15.cm]{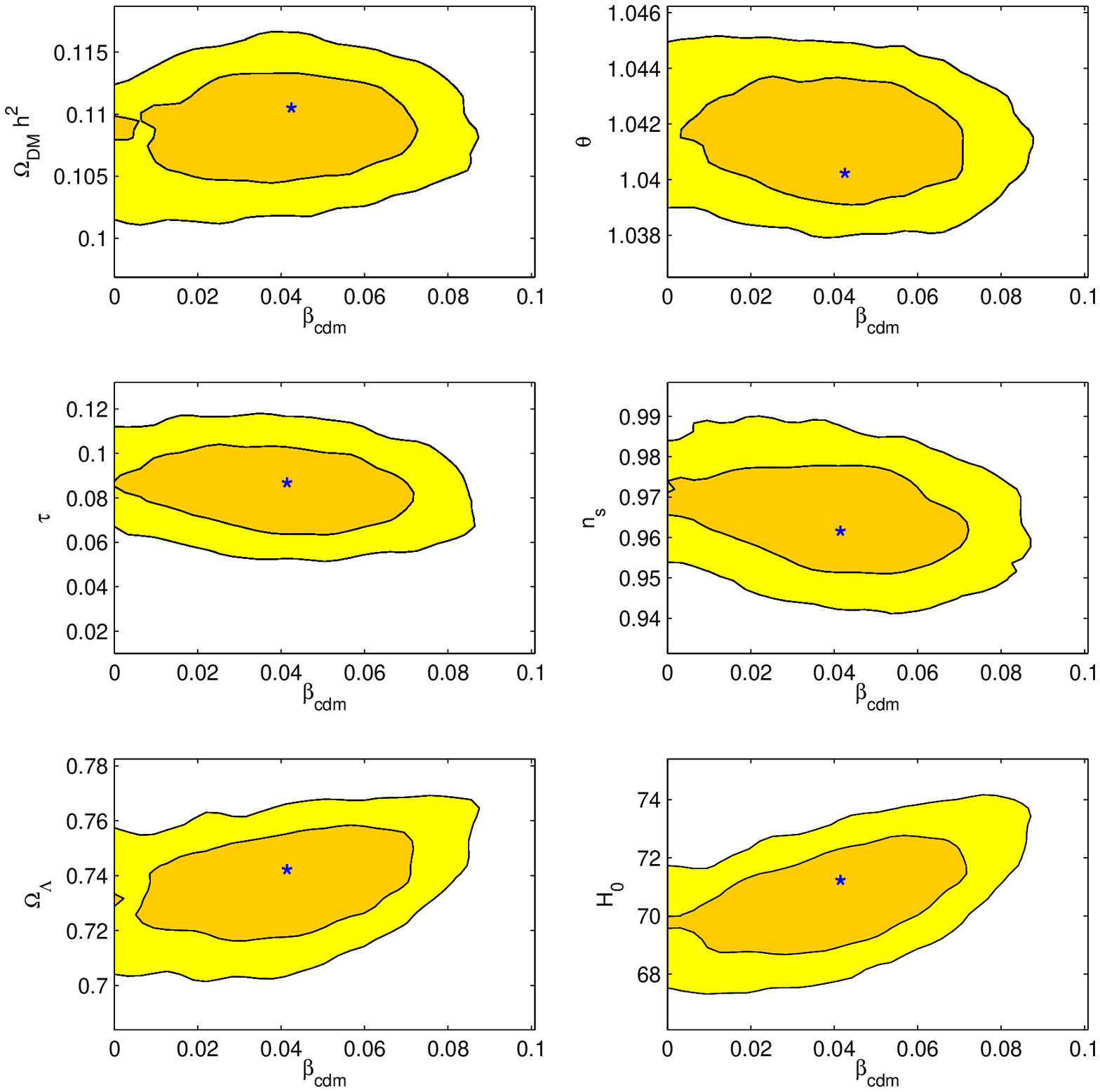}
      \caption{\small Likelihood contours for cosmological parameters when data from HST, BAO and Supernovae Ia are included. Blue asterisks mark best fit points for run $cq1hst$.}
 \label{fig:like_cont_hstv2}
 \end{figure*}

As discussed earlier on in this paper and in \cite{amendola_etal_2012}, the
coupling $\beta$ is degenerate with the Hubble parameter. In order to
investigate the effect of this degeneracy on the constraints from data, we did
another run ($cq1hst$) in which we combined the data used for $cq1$ (WMAP7 and
SPT) to also included baryon acoustic oscillations (BAO) \cite{percival_etal_2010}, Hubble Space Telescope constraints on $H_0$ (HST) \cite{riess_etal_2011} and Supernovae Ia (SNae) data \cite{union2_2010} as from COSMOMC (Aug 2011).
Likelihood contours for this case are shown in Fig.{\ref{fig:like_cont_hstv2}}.
Best fit values and errors are shown in Tab.\ref{tab:cq1mock_bestfit}, left
column. There seems to be an interesting preference for a non zero coupling,
though the values are clearly still compatible with zero at 1$\sigma$.
This peak comes mostly from a slight tension between the Hubble parameter HST result ($h=0.738\pm 0.024 $) and our
 WMAP7+SPT best fit for $\beta=0$ ($ h=0.685\pm 0.025 $ ). 
Notice however that even for $\beta=0$ we are not in an exact $\Lambda$CDM
since in our model $w$  is close, but not exactly equal, to -1. 
It is interesting then to test whether the  forecoming data from Planck can confirm or reject  this non zero coupling.
This we do in the next section.

 \normalsize

\subsection{Combining Planck and SPT mock data}
\begin{figure*}
 \centering
    \includegraphics[width=18.cm]{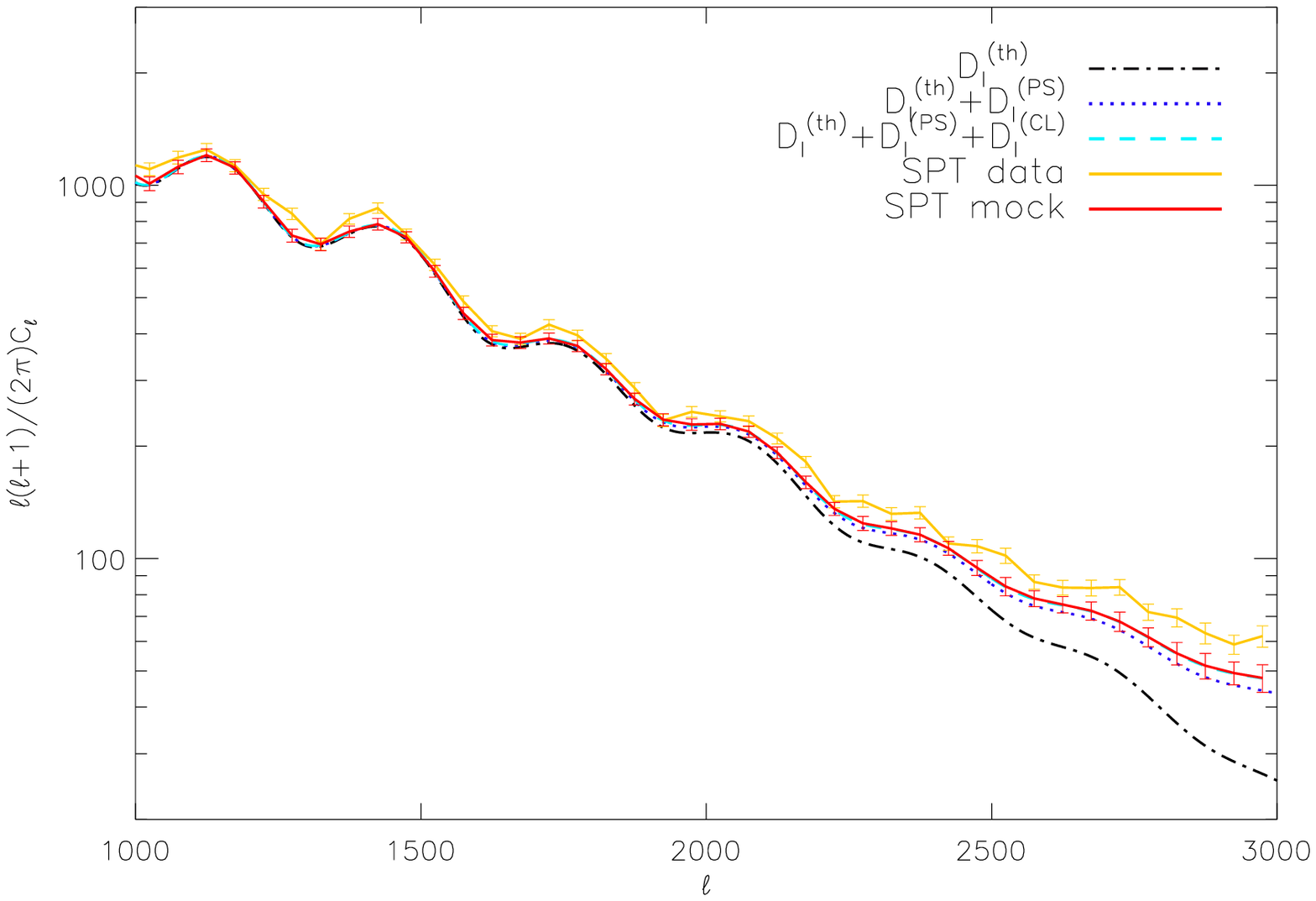}%
      \caption{\small SPT mock data as obtained from the fiducial $\Lambda$CDM spectrum.}
 \label{fig:mock}
 \end{figure*}

As a further analysis, we forecast the effect that Planck data would have on the
coupling parameter, when combined with the power measurement of $SPT$ (run
$cq1Pl$).
Since Planck data are not yet available, we produce a set of mock data
\cite{planck2011}. We have therefore implemented FutureCMB
\cite{perrotto_etal_2006_futurecmb} in our modified version of COSMOMC and,
using as fiducial power spectrum a
$\Lambda$CDM model
$D_l^{th}$, with $D_l = l(l+1)/2\pi  C_l$ (black dot-dashed in
Fig.\ref{fig:mock}).
We have then produced an SPT mock spectrum as follows:
\begin{description}
\item[ ]  we have added to the same fiducial model used to generate Planck mock
data, the effect of poisson sources (PS) (blue dotted) using the value for the
nuisance parameters ${{D_l}^{PS}}_{3000} = 18.1 \mu K^2$ with a dependence from
the momentum $\propto (l/3000)^2$  \cite{k11};
\item[ ] we have added to $D_{l}^{th}$ + PS the effect of clustered sources (CL)
(light blue dashed line), using the value for the nuisance parameters
${{D_l}^{CL}}_{3000} = 3.5 \mu K^2$ with a dependence from the momentum $\propto
(l/3000)^0.8$ for all $l$. This is not entirely correct as the dependence is
slightly different for $l > 1500$ and $l < 1500$ but differences are thought to
be small \cite{k11}; we obtain the $\tilde{D}_l \equiv D_l^{th} + PS + CL$. We
neglect here the effect of SZ. 
\item[ ] we have then convolved the $\tilde{D}_l$ for the SPT window functions
and created an SPT mock data with $D_l$ given by the convolved $\tilde{D}_l$ and
errors given by SPT data from \cite{k11}. The final SPT mock data is plotted in
Fig.\ref{fig:mock}) (red dotted line).
\end{description}

 \normalsize

\begin{center}
\begin{table}
\begin{tabular}{lllll}
\textbf{Best fit values} & \textbf{for coupled quintessence}\\
\hline
\begin{minipage}{30pt}
\flushleft
Parameter 
\\
\end{minipage} & 
\begin{minipage}{150pt}
\flushleft
 \emph{cq1hst} (baseline + $\beta$+$\sigma$)\\ real data
\end{minipage} &  
\begin{minipage}{150pt}
\flushleft
$cq1Pl$  \\ forecasts
\end{minipage} &
\\
\\
\hline
$  \Omega_b h^2  $ & $0.022^{+00044}_{-0.00040}$           & $0.023^{+0.0001}_{-0.00024} $		\\
$ \Omega_{c} h^2  $ & $0.11^{+0.0022}_{-0.0039}$                                  & $0.11^{+0.0013}_{-0.0012}$ \\
      $ \theta_s$ & $1.04^{+0.0016}_{-0.0014} $                    & $1.05^{+0.00008}_{-0.0006}$ \\
   $ \tau$ & $0.084^{+0.0061}_{-0.0067} $                             & $0.090^{+0.00011}_{-0.0039}$\\   
   $ n_s$ & $0.96^{+0.010}_{-0.0086} $                                                     & $0.98^{+0.0024}_{-0.0062} $ \\
   $w$ &  $-0.90^{+0.10}_{-0.098} $                                                                                 & $-1.0^{+0.1}_{-0.003}$ \\  
   $\beta$ &  $ 0.041^{+0.0092}_{-0.041}$                          & $ 0.0035^{+0.0089}_{-0.0035} $\\  
   $\beta$ &  $ < 0.050 (0.074)$                          & $ < 0.012 (0.030) $\\  
    $ \sigma$ & $0.19^{+0.31}_{-0.059}$                                              & $0.48^{+0.015}_{-0.11}$ \\
   $ \Omega_{de}$ & $0.74^{+0.012}_{-0.015} $                                  & $0.73^{+0.019}_{-0.000056} $ \\
   $ Age/Gyr$ & $13.7^{+0.082}_{-0.093} $                                              & $13.6^{+0.005}_{-0.08}$ \\
   $ z_{re}$ & $10.3^{+1.1}_{-1.2} $                          & $10.6^{+0.2}_{-0.5} $ \\
   $ H_0$ & $71.0^{+1.0}_{1.7}$                             & $ 69.6^{+2.4}_{-0.13} $  \\ 
         $ D_{3000}^{SZ}$ & $5.0^{+1.2}_{-5.0} $                    & $2.1^{+2.8}_{-1.1} $ \\
                  $ D_{3000}^{PS}$ & $20.6^{+1.9}_{-3.4} $                    & $15.6^{+2.6}_{-2.3} $ \\
                           $ D_{3000}^{CL}$ & $5.1^{+3.0}_{-1.5} $                    & $2.9^{+1.0}_{-2.8} $ \\
   -$Log$(Like) &$ 4024$                      							& $660 $  \\ 
\hline
\hline
\end{tabular}
\caption{Best fit values and 1-$\sigma$ errors comparing runs  \emph{cq1hst} and
mock $cq1Pl$ including SPT + PLANCK, plus HST, BAO, SN data. For $\beta$ we also
write in brackets the value of the 2$\sigma$ marginalized error.}
 \label{tab:cq1mock_bestfit}
\end{table}
\par\end{center}
\normalsize

\begin{figure*}
 \centering
    \includegraphics[height=12.cm]{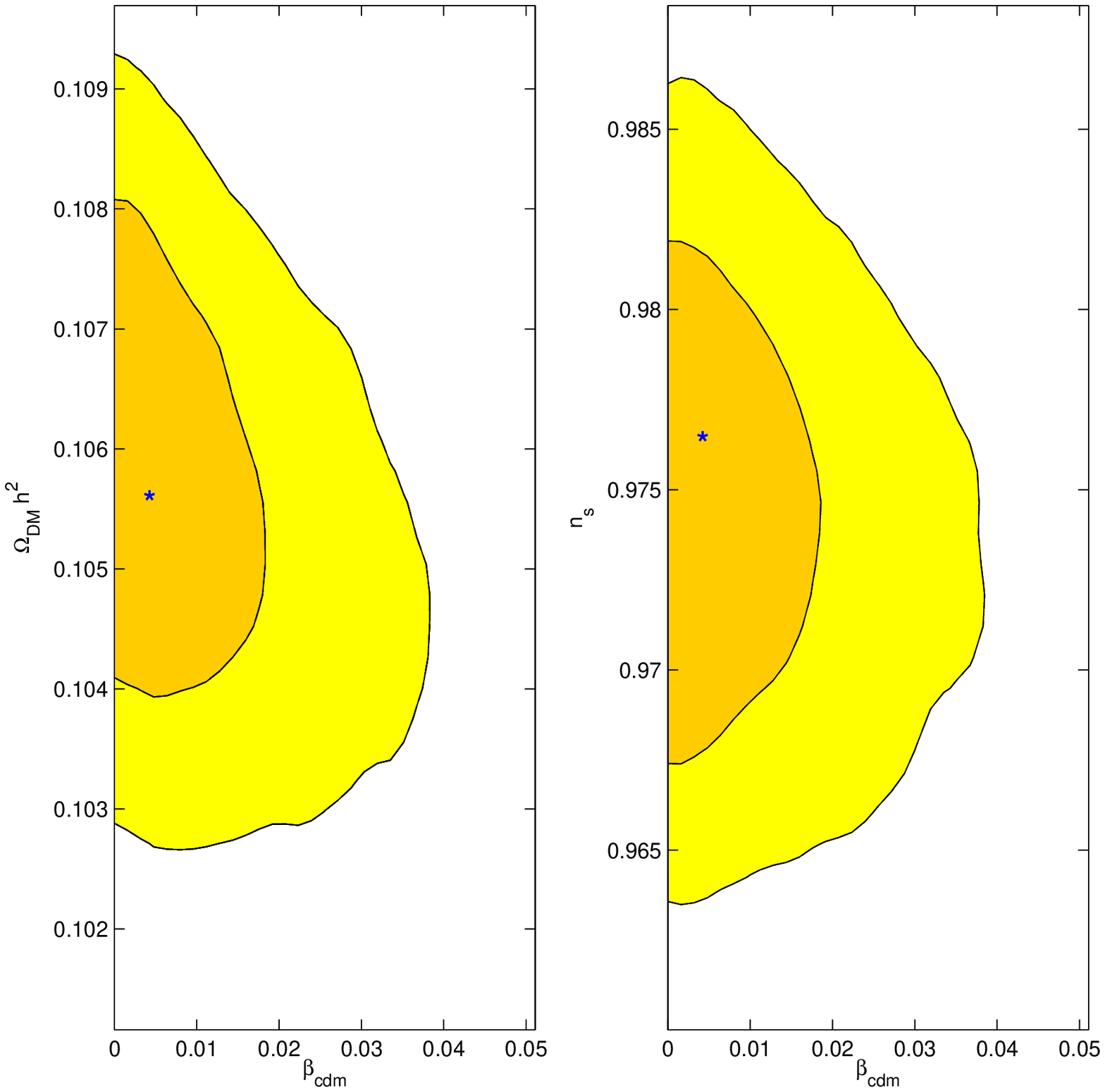}%
      \caption{\small Confidence contours for Planck + SPT mock. Blue asterisks mark best fit points for run $cq1Pl$.}
 \label{fig:like_cont_curvature_mock}
 \end{figure*}

For this case, we also include BAO, HST and Supernovae Ia data as from Cosmomc (August 2011 version), 
to break the degeneracy between the coupling and the Hubble parameter. Some representative
2D confidence regions are in Fig. \ref{fig:like_cont_curvature_mock}.
We then find that $\beta < 0.012 (0.030)$; though stronger than WMAP7+SPT, this is still a pessimistic bound,
since it includes Planck but still considers SPT data with errors released by
\cite{k11}; by the time Planck data will be available, better SPT (or ACT) data
may have been released. In Table \ref{tab:cq1mock_bestfit} we report the best
fit values, together with the left and right errors at $68 \%$ and $95\%$ CL
around the best fit. The left column refers to run $cq1hst$, done with real data
(WMAP+SPT+HST+BAO+SNae); the right column shows the forecasted values for
Planck+SPT mock data, plus SNae, HST and BAO.

 \normalsize

\begin{figure*}
 \centering
    \includegraphics[width=10.cm]{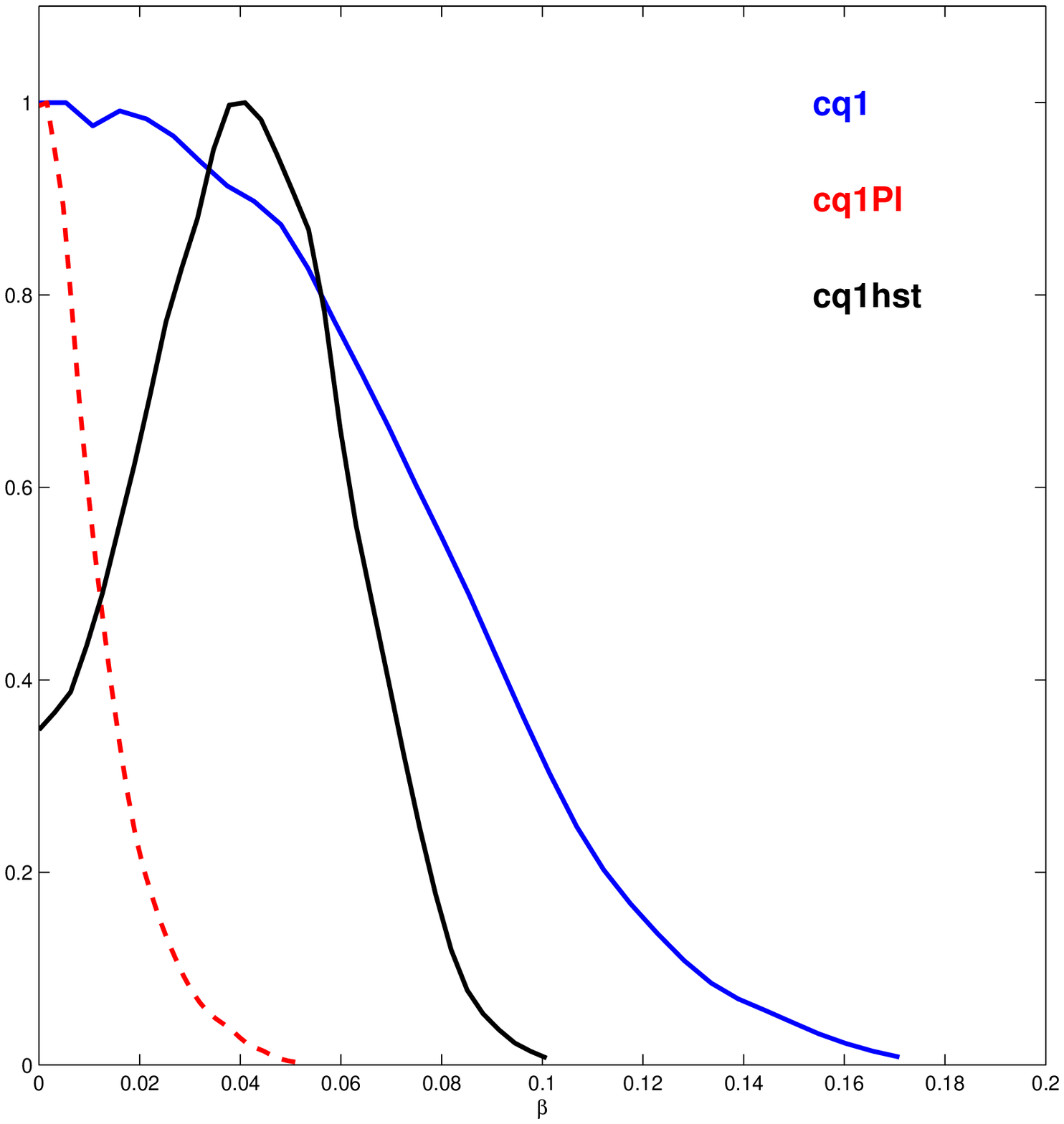}%
      \caption{\small 1D likelihood for the coupling $\beta$ from runs $cq1$, $cq1hst$, $cq1Pl$. We recall that $cq1$ and $cq1hst$ are based on real data while $cq1Pl$ estimates the forecasted constraints from mock data, around a fiducial $\Lambda$CDM cosmology.}
 \label{fig:finalbeta}
 \end{figure*}

We finally plot in Fig. \ref{fig:finalbeta} a comparison between current data and
future observations,  marginalizing over all parameters except $\beta$.
It is clear  that Planck data will reach a precision sufficient  to tell whether
the peak in $\beta$ is a real detection or just a fluke.

 \normalsize

\section{Conclusions} \label{conclusions}
We have considered the possibility that the evolution of dark matter and dark energy might be
connected by a constant coupling, of the type illustrated in
\cite{amendola_2000, pettorino_baccigalupi_2008}. We have used current CMB data
from WMAP7 and SPT to constrain the coupling parameter $\beta$. We find that
$\beta$ is constrained to be less than 0.063 (0.11) at $68\%$ ($95\%$) C.L. when SPT data are
included, with respect to $\beta < 0.078 (0.14)$ coming from WMAP7 only. 
We have done a number of tests to check whether this bound depends on the degeneracy with other parameters (lensing, curvature, massive neutrinos, $N_{eff},$ HST/BAO/SNae data).
If the effective number of relativistic degrees of freedom $N_{eff}$ is allowed to
vary, no much gain is obtained on $\beta$, which still needs to be $\beta <
0.074 (0.12)$.
We have further considered the effect of CMB-lensing, both with a run which
includes no lensing and by marginalizing over $A_L$, a parameter which encodes
the rescaling of the lensing power spectrum. $A_L$ is slightly degenerate with
$n_s, \Omega_{DM}h^2$ which in turn are degenerate with $\beta$, though no
direct degeneracy is seen between $\beta$ and $A_L$.
If the assumption of a flat universe is released, constraints on $\beta$ weaken
back almost to the level of constraints given by WMAP only (flat universe), with $\beta <
0.071 (0.13)$. Degeneracy with massive neutrinos widens the coupling constraints to be $\beta < 0.084 (0.14)$ when we marginalize over the fraction of massive neutrino species $f_{\nu}$.
We conclude that the bound on $\beta$ from current data is already strong enough to be quite stable with respect to a better knowledge of other parameters and to all cases considered.

When WMAP+SPT are considered (run $cq1$), the best fit value for $\beta$, though
still fully compatible with zero, has a best fit of $\beta =
0.012^{+0.050}_{-0.012}$. It is interesting to see that when we allow for an
effective number of relativistic degrees of freedom, marginalizing over
$N_{eff}$ the coupling from WMAP7+SPT data increases to a best fit value of
$\beta \sim 0.03$. A larger value of $N_{eff}$ favors larger couplings between
dark matter and dark energy and values of the spectral index closer to $1$.
Including SPT data does not improve significantly constraints on the coupling
$\beta$. 
Inclusion of additional priors from HST, BAO and SNae moves the best fit to $\beta=0.041$, again still compatible
with zero at 1$\sigma$.
We forecast that the inclusion of Planck data will be able to
pin down the coupling to about $1\%$ and therefore detect whether the small non-zero
coupling present in current data is washed away  with more data.

\begin{acknowledgments}
V.P. is supported by the Marie Curie IEF, Project DEMO - Dark Energy Models and Observations. Support was given to V.P. and C.B. by the Italian Space Agency through the ASI contracts Euclid-IC (I/031/10/0). V.P. thanks Joanna Dunkley for useful discussion and advices. We thank Christof Wetterich for helpful comments.
C.B. also acknowledges support from the PD51 INFN initiative. 
L.A. acknowledges support from DFG - Deutsche Forschungsgemeinschaft through the project TRR33 "The Dark Universe".
\end{acknowledgments}
\bibliography{pettorino_bibliography}

\end{document}